\documentclass[journal,hideappendix]{vgtc}                     

\onlineid{0}

\vgtccategory{Research}

\vgtcpapertype{evaluation}

\title{Quantifying Emotional Responses to Immutable Data Characteristics and Designer Choices in Data Visualizations}

\author{%
  \authororcid{Carter Blair}{0009-0001-9472-4071},
  Xiyao Wang, and 
  \authororcid{Charles Perin}{0000-0002-7324-9363}
}

\authorfooter{
  \item
  	Carter Blair is with the University of Waterloo.
  	E-mail: cblair@uwaterloo.ca
  \item
  	Xiyao Wang is with the University of Victoria and TU Delft. 
        E-mail: xiyao.wang23@gmail.com

  \item Charles Perin is with the University of Victoria.
        E-mail: cperin@uvic.ca
}

\abstract{%
  Emotion is an important factor to consider when designing visualizations as it can impact the amount of trust viewers place in a visualization, how well they can retrieve information and understand the underlying data, and how much they engage with or connect to a visualization. We conducted five crowdsourced experiments to quantify the effects of color, chart type, data trend, data variability and data density on emotion (measured through self-reported arousal and valence). Results from our experiments show that there are multiple design elements which influence the emotion induced by a visualization and, more surprisingly, that certain data characteristics influence the emotion of viewers even when the data has no meaning. In light of these findings, we offer guidelines on how to use color, scale, and chart type to counterbalance and emphasize the emotional impact of immutable data characteristics.
}


\teaser{
  \centering
      \includegraphics[alt={Infographic detailing factors in data visualization across six categories: Method: States that SAM (Self-Assessment Manikin) scale and unstructured study methods are valid as they allowed to replicate results from Bartram et al (7) (S1, H1). Color: Notes that grayscale color palette is impactful, but its effect is the most negative of all palettes tested (S1, H2). It also notes that studying color outside of visualization is transferable to visualization (S2, H1). Data Trend: Indicates positive data trends induce higher valence and arousal than neutral trends, which in turn induce higher valence and arousal than negative trends (S2, H2 \& 3). Data Variance: Shows higher data variance induces higher arousal (S4, H1). Data Density: States higher data density induces higher arousal (S5, H1). Chart Type: Lists several comparisons: Charts that are good at showing trend like line chart and scatterplot induce more emotion than those that are not, like bar chart (S3, H2 \& 3). Line charts induce more emotion than scatter plots. Data trend has a much stronger effect than chart type on emotion (S3, H6). Charts with round features like smoothed line charts induce higher valence than charts with sharp features like jagged line charts (S3, H5). Line charts induce higher arousal than scatterplots. Each category uses pastel color coding and contains citations to specific studies.}, width=\linewidth]{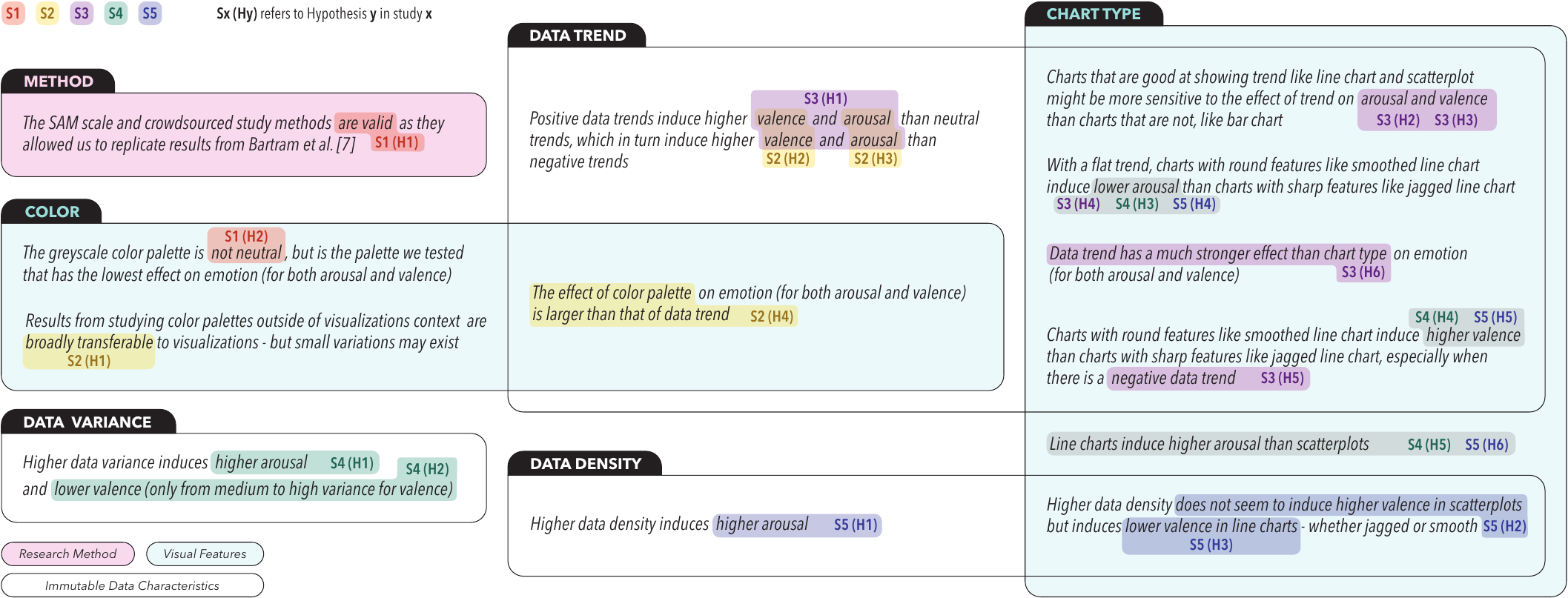}
  \caption{Summary of results from the five studies S1--S5.}
  \label{fig:summary-results}
}

\graphicspath{{./figures/}{./figures/s5/}{./figures/s4/}{./figures/s3/}{./figures/s1/}{./figures/s2/}{./}}

\usepackage{mathptmx}                  
\usepackage{graphicx}
\usepackage[table,svgnames]{xcolor}
\usepackage{xspace}
\usepackage{tikz}
\usetikzlibrary{arrows.meta,chains,positioning,fit,calc,shapes.geometric}
\usepackage{colortbl}
\usepackage{array}
\usepackage{tabularx}
\newcolumntype{M}[1]{>{\centering\arraybackslash}m{#1}}

\newcommand{\eg}{e.\,g.,\xspace}

\newcommand{\etal}{et~al.\xspace}

\newcommand{\errorbardescrip}{}
\newcommand{\inlinesection}[1]{\noindent\textbf{\emph{\underline{#1.}}}\xspace}

\usepackage{soul,transparent}

\makeatletter
 \def\SOUL@hlpreamble{%
 \setul{}{2ex}
 \let\SOUL@stcolor\SOUL@hlcolor
 \SOUL@stpreamble
 }
\makeatother

\definecolor{arousal}{HTML}{EFD0E0}
\definecolor{valence}{HTML}{BEE2EA}
\definecolor{guideline}{HTML}{BEEAD6}
\definecolor{amendment}{HTML}{FFFF00}
\newcommand{\arousalCol}[1]{\sethlcolor{arousal}\hl{\textit{#1}}\xspace}
\newcommand{\valenceCol}[1]{\sethlcolor{valence}\hl{\textit{#1}}\xspace}
\newcommand{\guidelineCol}[1]{\sethlcolor{guideline}\hl{\textit{#1}}\xspace}

\newcommand{\delete}[1]{\xspace}
\newcommand{\add}[1]{#1\xspace}

\begin{document}


\firstsection{Introduction}

\maketitle


Emotion is an important factor to consider when designing data visualizations. 
Viewer’s emotion can impact the extent to which they engage with or connect to a visualization~\cite{cairo2013emotional,kennedy2018feeling,Lan:2022:NEP} as well as their trust in~\cite{lin2021fooled}, and understanding of~\cite{harrison2012exploring}, the data. 
\add{Emotional response matters in many decision-making and communication scenarios: a healthcare provider might want a patient to stay calm when presented with medical information, a media might aim to arouse viewers, or a business might aim to get investors excited about their financial results.}

The field of psychology has established that visual stimuli can induce emotion~\cite{royet2000emotional,hermans1994affective}. 
It has been found that this is also true for both static~\cite{Bartram:2017:ACV} and dynamic~\cite{kennedy2018feeling,lan2022chart,Lockyer:2012:AMT} data visualizations. 
However, empirical evaluations are missing to quantify the effects of visualization features other than color~\cite{Bartram:2017:ACV} on emotion. 
This is an important gap to address so that visualization designers can make informed design choices to counterbalance and enhance the emotional impact arising from immutable data characteristics in visualizations.

We show that \textbf{chart type}, \textbf{data trend}, \textbf{data variance}, and \textbf{data density} all influence the emotion induced by a visualization---in addition to \textbf{color}~\cite{Bartram:2017:ACV}. 
We quantify through five studies the effects of color (S1 and S2), chart type (S3, S4, and S5), data trend (S2 and S3), data variance (S4), and data density (S5) on emotion (measured through arousal and valence ratings using the Self-Assessment Manikin scale~\cite{Lang1980}).

Remarkably, this work is the first to show that \textbf{\textit{even if the data has no semantic meaning, certain data characteristics influence the emotion of viewers}}. \Cref{fig:summary-results} summarizes the results from the five studies. 
It shows that two fundamental designer choices (color and chart type) and three fundamental immutable data characteristics (data trend, variance, and density) influence viewers' emotions regardless of a visualization's semantic meaning. 
From these results, we provide guidelines on how to manipulate color, scale, and chart type to counterbalance or emphasize the emotional impact of immutable data characteristics. This is a starting point for fully understanding the effect of visualization features on viewers' emotions.




\section{Related Work}
Work in psychology and visualization \add{sheds light on potential effects of designer choices and immutable data characteristics on emotions}. 

\subsection{The role of emotion in visualization}
Emotional response should be considered to design and assess the value of visualizations~\cite{d2016feminist,wang2019emotional,lan2023affective} for several reasons:
emotional priming with positive emotion-inducing stimuli increases the accuracy of information extraction from visualizations~\cite{harrison2012exploring};
negative emotions have controversial effects and can sometimes increase long-term recall~\cite{Lan:2022:NEP}; 
emotion can affect engagement with data~\cite{kennedy2018feeling};
and there is a relationship between color, perceived beauty, and the degree to which a visualization is trusted by viewers~\cite{lin2021fooled}. 
In data storytelling also, creating narratives that induce emotion is an important consideration~\cite{10.5555/3235128} to help viewers engage with and relate to the information~\cite{cairo2013emotional}. 

Together, these factors emphasize the significance of investigating the elements of visualizations that impact emotion and their extent.

\subsection{Measuring emotion}
\label{sec:rw-measure}

Measures of emotion can be behavioral, physiological, or experiential~\cite{mauss2009measures, bradley2000measuring}.
Behavioral measures quantify overt acts and secondary behaviors such as voice pitch (\eg~\cite{bachorowski1999vocal}) or facial behavior (\eg~\cite{russell1994there}).
Physiological measures (\eg~\cite{riche2010beyond}) focus on physiological reactions such as changes in brain states or skin conductance. 
Experiential measures (\eg~\cite{lyra2016infographics}) 
quantify self-reported subjective experiences. 
\textit{Behavioral} and \textit{physiological} measures \add{require moderation because they} 
rely on 
dedicated equipment such as audio and video recordings and physiological sensors. 
We \add{used} \textit{experiential} measures 
\add{because} 
\add{i)} self-assessment can be administered as a 
survey question, \add{and ii)} they are reliable to measure arousal and valence in crowdsourced studies~\cite{baveye2014crowdsourced}. 

Emotion is commonly characterized through three factors: arousal, valence, and dominance~\cite{bradley1994measuring}. 
Given a stimulus, arousal corresponds to the intensity of emotion it induces, valence to its pleasantness and dominance to its perceived power~\cite{warriner2013norms}. 
Because arousal and valence are more significant and discriminatory~\cite{bradley1994measuring, mehrabian1974approach} and more reliable~\cite{bradley1994measuring, mauss2009measures} measures than dominance, we used the SAM rating system~\cite{Lang1980} to measure arousal and valence on nine-point scales (see \Cref{fig:scales}).

\subsection{The effect of color on emotions}
Color has been the most studied feature when it comes to emotion. 
Its role has been largely explored in psychology. 
It has been shown, for example, that the printed paper color has an influence on emotions when reading~\cite{Weller:1988:ECQ};
that saturated and bright colors are associated with higher arousal and valence and that hue has an effect on both measures~\cite{wilms2018color};
that principle hues (such as red, yellow, and green) elicit more positive responses than intermediate hues (such as yellow-red and blue-green), which in turn elicit more positive responses than achromatic colors such as grey and black~\cite{kaya2004relationship}; and
that bright colors are associated with high valence and saturated colors with high arousal~\cite{valdez1994effects}.

The role of color on emotion has also received attention in visualization; unsurprisingly, given the importance of color in visualization design (\eg~\cite{heer2012color,lin2013selecting,moreland2009diverging,rogowitz1998data,silva2011using,szafir2017modeling,ware2019information}). 
Bartram \etal~\cite{Bartram:2017:ACV}, building on results from psychology, studied the relationship between color in bar charts and emotion, from which they designed a series of palettes that convey different emotions.
Anderson \etal~\cite{Cary:2022:ACV} further validated the emotional impact of varying color palettes when interpreting categorical maps.
A recent study on arousal-related design features also highlighted the affective influence of color on visualizations~\cite{lan2022chart}.

\subsection{What other features could affect emotions?}
\label{sec:rw-other}
Other features than color have an effect on emotion. 
Motion, for example, affects emotion~\cite{Bartram:2010:WMM, Lockyer:2012:AMT}; and for a static visualization, the trajectory of the data can provide a sense of motion.
Shape can also affect emotion, given that humans tend to prefer rounded objects over sharp objects~\cite{bar2006humans}, with sharp objects having high arousal and low valence~\cite{aryani2020affective}.
Leder \etal~\cite{leder2011emotional} distinguished positive (like cake) and negative (like snake) stimuli and found that people's preference towards curved contours over sharp ones held for positive or neutral stimuli, but no difference was identified for negative stimuli. 
Various other studies in psychology have shown that people's preferences vary with different shape properties, such as symmetry~\cite{TINIO2009241}, size~\cite{Silvera:2002:bib}, and complexity~\cite{Phillips2010FechnersAR}.

The chart type influences the perception of trend and determines the shape of the visualization, affecting users' preference~\cite{Moere:2012:EES} and memorability~\cite{borkin2013makes}. 
Interestingly, motion and shape can be manipulated by a visualization designer (\eg using motion to encode data and using circles vs. squares to represent data points in a scatterplot), but they are also influenced by immutable data characteristics such as trend, variance and density, that create the direction of motion. 


Based on the above, and acknowledging that many features and data characteristics could be studied, we identified a first set of visualization aspects to study. In addition to color, we include chart type, data trend, data variance, and data density. We chose these because i) they are potentially emotion-inducing features, ii) together they allow us to investigate both features that are due to immutable data characteristics that designers cannot manipulate and features that visualization designers can manipulate, and iii) we lack empirical data about the relationship between these characteristics and features, and emotion.

\begin{figure}[!t]
\centering
  \includegraphics[alt={The image displays the Self-Assessment Manikin (SAM) scales for measuring arousal and valence in studies. It consists of two rows, each with nine positions numbered 1 through 9. The top row, labeled "AROUSAL" with a pink background, shows five pictograms of increasing excitement. It starts with a calm figure, progresses to more active poses, and ends with an extremely aroused figure surrounded by stars. The bottom row, labeled "VALENCE" with a light blue background, depicts five faces ranging from frowning to neutral to broadly smiling. Both scales use stylized human-like figures to represent varying emotional states, with empty spaces between the numbered positions.}, width=.9\columnwidth]{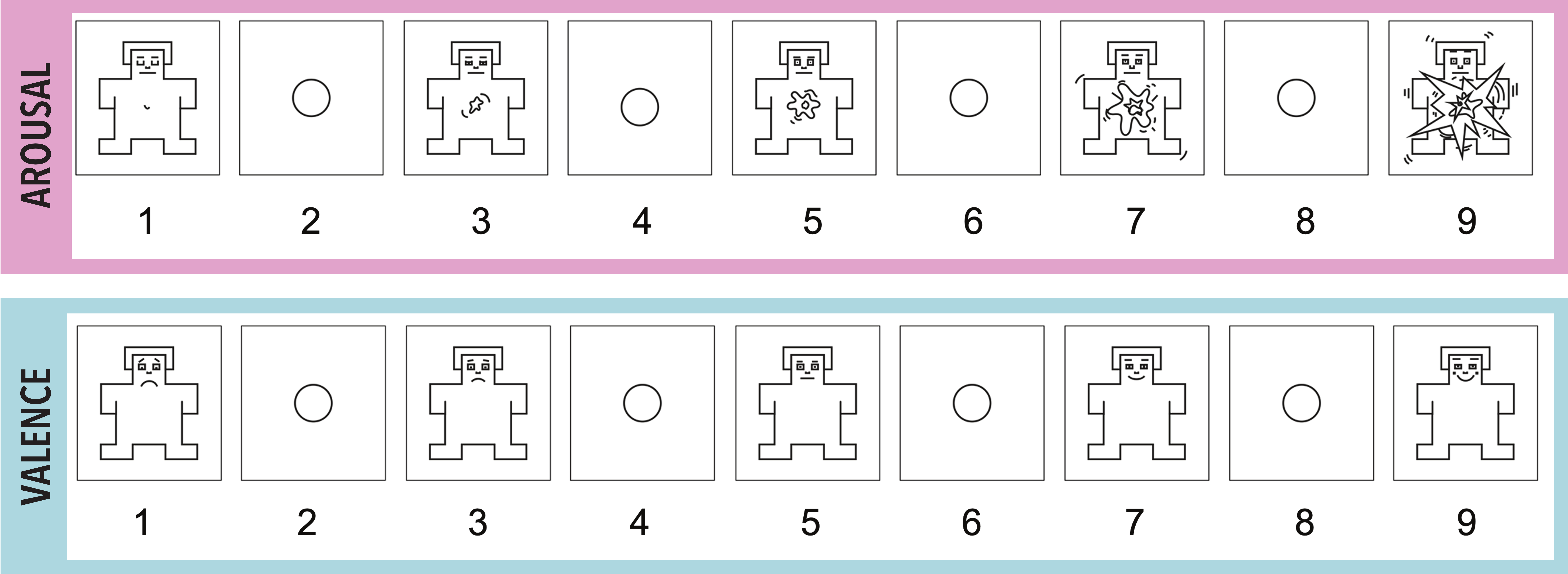}
  \caption{SAM scales used in our studies (figures from~\cite{betella2016affective}).}
  \label{fig:scales}
  \vspace{-1.5em}
\end{figure}

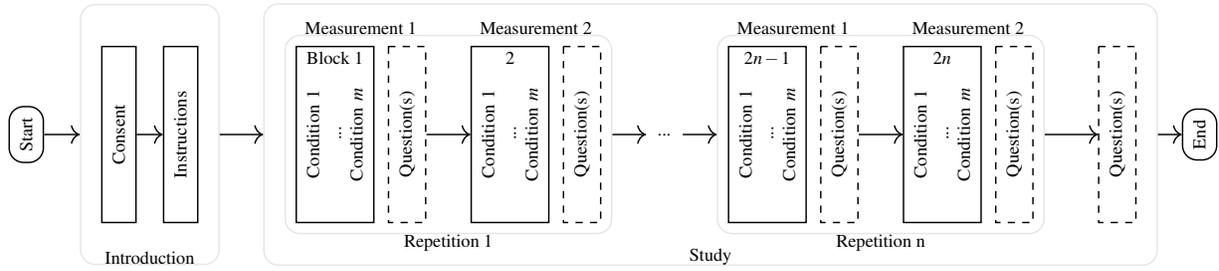
\begin{figure*}[!t]
    \centering
    \resizebox{0.9\textwidth}{!}{
    \begin{tikzpicture}[
    every node/.append style={font=\tiny},
    startstop/.style = {rectangle, rounded corners,text centered, draw=black},
    startstop_vertical/.style = {rectangle, rounded corners,text centered, draw=black,rotate=90,anchor=north},
  block/.style = {draw, align=center,text width=1.5cm},
  block_vertical/.style = {draw, align=center,text width=1.5cm,rotate=90,anchor=north},
  line/.style={-latex},
  dotted_block/.style={draw=black!10!white, rectangle, rounded corners},
block_vertical_dash/.style={text width=1.5cm,  align=center, draw, dash pattern=on 2pt off 2pt, rectangle,rotate=90,anchor=north}
                    ]
    \begin{scope}
\node (s1) [startstop_vertical] {Start};
\node (i1) [block_vertical,below of = s1, node distance = 9mm]  {Consent};
\node (i2) [block_vertical, below of = i1, node distance = 6mm]  {Instructions};

\node (sb1) [block_vertical, below of = i2, node distance = 15mm]  {Condition 1 \\ ... \\ Condition $m$ };
\node at (sb1.east) [below, inner sep=0.5mm]  {Block 1};
\node (q1) [block_vertical_dash, below of = sb1, node distance = 7mm]  {Question(s)};
\node (sb2) [block_vertical, below of = q1]  {Condition 1 \\ ... \\ Condition $m$ };
\node at (sb2.east) [below, inner sep=0.5mm]  {2};
\node (q2) [block_vertical_dash, below of = sb2, node distance = 7mm]  {Question(s)};
\node (sbi) [align=center, right of = q2, node distance = 8mm]  {...};
\node (sb3) [block_vertical, below of = sbi, node distance = 10mm]  {Condition 1 \\ ... \\ Condition $m$ };
\node at (sb3.east) [below, inner sep=0.5mm]  {$2n-1$};
\node (q3) [block_vertical_dash, below of = sb3, node distance = 7mm]  {Question(s)};
\node (sb4) [block_vertical, below of = q3]  {Condition 1 \\ ... \\ Condition $m$ };
\node at (sb4.east) [below, inner sep=0.5mm]  {$2n$};
\node (q4) [block_vertical_dash, below of = sb4, node distance = 7mm]  {Question(s)};
\node (q5) [block_vertical_dash, below of = q4]  {Question(s)};

\node(measure1)[inner ysep=1mm,inner xsep=1mm,fit=(sb1)(q1)]{};
\node at (measure1.north) [above, inner sep=0mm]  {Measurement 1};
\node(measure2)[inner ysep=1mm,inner xsep=1mm,fit=(sb2)(q2)]{};
\node at (measure2.north) [above, inner sep=0mm]  {Measurement 2};
\node(measure3)[inner ysep=1mm,inner xsep=1mm,fit=(sb3)(q3)]{};
\node at (measure3.north) [above, inner sep=0mm]  {Measurement 1};
\node(measure4)[inner ysep=1mm,inner xsep=1mm,fit=(sb4)(q4)]{};
\node at (measure4.north) [above, inner sep=0mm]  {Measurement 2};

\node(rep1)[dotted_block, inner ysep=1mm,inner xsep=1mm,fit=(sb1)(q2)]{};
\node at (rep1.south) [below, inner sep=0mm]  {Repetition 1};
\node(rep2)[dotted_block, inner ysep=1mm,inner xsep=1mm,fit=(sb3)(q4)]{};
\node at (rep2.south) [below, inner sep=0mm]  {Repetition n};

\node(block1)[dotted_block, inner ysep=4mm,inner xsep=2mm,fit=(i1)(i2)]{};  
\node at (block1.south) [above, inner ysep=0mm]  {Introduction};
\node(block2)[dotted_block, inner ysep=3mm,inner xsep=2mm,fit=(rep1)(q5)]{};  
\node at (block2.south) [above, inner sep=0mm]  {Study};

\node (e1) [startstop_vertical, below of = q5, node distance = 8mm] {End};

\draw[->](s1) -- (block1);
\draw[->](i1) -- (i2);
\draw[->](rep1) -- (sbi);
\draw[->](q1) -- (sb2);
\draw[->](sbi) -- (rep2);
\draw[->](q3) -- (sb4);
\draw[->](rep2) -- (q5);
\draw[->](block1) -- (block2);
\draw[->](block2) -- (e1);
    \end{scope}
    \end{tikzpicture}
    }
    \caption{Workflow for the five studies. Dashed blocks are optional question(s) asked depending on each study design. Measurements 1 and 2 are arousal or valence, depending on the counter-balanced order.}
    \label{fig:workflow}
    \vspace{-1em}
\end{figure*}


\section{Study Rationale and Methodology}
\label{sec:method}

The aim of this research is to examine the relationship between emotion and three immutable data characteristics (data trend, data variance and data density), and two visual features of visualizations (color and chart type). 
To achieve this, we conducted five crowdsourced studies (S1--S5, approved by our institutional Ethics board) that asked participants to rate their self-perceived arousal and valence in response to visualizations. 
We study color (S1); color and data trends (S2); data trends and chart types (S3); data variance and chart types (S4); and data density and chart types (S5).
\add{We chose those factors due to their ubiquity in visualization, and we studied at most two factors per study to keep the participant time requirements reasonable. Studies 2, 3, 4, and 5 each had two independent variables---one a design choice and the other a data characteristic.}
Next, we describe the methodology common to all studies (the specificity of each is provided in their respective sections).

\inlinesection{Measurements}
We used the SAM rating system~\cite{Lang1980} to measure arousal and valence on nine-point scales (see \Cref{fig:scales} and \Cref{sec:rw-measure}). 
In the remainder of the paper, 
\arousalCol{$A$} refers to the arousal ratings for $A$ and \valenceCol{$A$} to the valence ratings for $A$.
\arousalCol{$A > B$} means that the arousal for $A$ is larger than for $B$, and 
\valenceCol{$A > B$} that the valence for $A$ is larger than for $B$.

\inlinesection{Design} 
Studies are within-subject designs in which all participants completed the same tasks for all conditions. 
The studies are divided into measurement blocks, each containing all conditions for one measurement. 
The number of blocks depends on the number of repetitions and is counter-balanced. 
The order of conditions within each block is randomized. 
The studies were created with both \textit{SurveyMonkey} and \textit{Qualtrics} and were deployed through the \textit{Prolific} crowdsourcing platform.
\Cref{fig:workflow} shows the structure of the studies.

\inlinesection{Procedure}
Each participant completed the study online without supervision. 
After selecting the study on Prolific, they were redirected to the study survey on SurveyMonkey/Qualtrics.
After participants had read the consent form and given consent to participate, they were given illustrated explanations about the rating scales (arousal and valence) and instructions on how to rate their emotions with these scales.
Then, they answered each question one by one -- they only saw the next question after the current one was confirmed, and they were not allowed to go back and change their answer.

\inlinesection{Participants} 
We recruited participants online through Prolific. 
They had to be over 18 years old, have normal or corrected to normal vision without color-deficiency, and have a Prolific approval rating over 99\% (to ensure quality responses). 
They were compensated at least minimum hourly wage based on the estimated duration of each study. 
We excluded participants who spent a total time far shorter than the estimated study time or who input the same answers for most conditions. 
We provide study participant demographics in supplemental materials.

\inlinesection{Analysis}
\add{Each study (aside from S1) had two independent variables $v_1$ and $v_2$ and two measures.
To assess the effects of $v_1$ and $v_2$, we used a 3-step procedure: for each participant $p$ and measure $m$,
1) Averaged $p$'s responses to all repetitions of a given \{$p$, condition, $m$\}, using the mean;
2) Calculated aggregate values for each level of $v_1$ and $v_2$ by averaging across all levels of the other variable. 
3) Generated confidence intervals (CIs) by either:
a) Bootstrapping aggregate values from all participants, or
b) Bootstrapping differences between levels from all participants.
There are two exceptions to this process: 1) in S1, there was only one variable, so we skipped step 2; 2) for hypotheses that concern the variance induced by an independent variable (S2(H4) and S3(H6)), we used standard deviation, as explained in \Cref{sec:s2:results}.}
We interpret the results using effect sizes and CIs rather than $p-$values, following recommendations for statistical practices in HCI and visualization (\eg~\cite{Cumming:2014:Stat, Dragicevic:2016:Stat}). We report the mean and 95\% CI values for each result. 


\section{S1 -- Color Palettes}
\label{sec:S1}
Study 1 investigates people's emotional responses to five color palettes. 
The first goal was to determine an emotionally neutral color palette that we could use in further experiments to isolate the emotional impact of other aspects of visualizations.
The second goal was to replicate the results from the study on affective colors~\cite{Bartram:2017:ACV} with the SAM rating scale.

\subsection{S1 -- Method}

\inlinesection{Design}
This study had five conditions (the five color palettes). 
Four come from Bartram et al.~\cite{Bartram:2017:ACV}'s work on affective color in visualization: the exciting ($P_{exc}$), positive ($P_{pos}$), negative ($P_{neg}$) and calm ($P_{cal}$) palettes, and we added a grey color palette ($P_{gre}$).
With three repetitions, this study consisted of six blocks \add{(three alternating arousal and valence blocks)}. 
Each block included \add{one instance of each of the five palettes}.

\begin{figure}[!t]
 \begin{minipage}[t]{\columnwidth}
 \centering
  \noindent
  \includegraphics[alt={This figure presents the color palettes utilized in Study 1, categorized by their associated emotional tones. The "Exciting" palette includes bright colors such as green, orange, and pink. The "Positive" palette features shades like yellow and light green. The "Negative" palette consists of darker hues including brown and maroon. The "Calm" palette contains pastel colors like peach and light blue. Lastly, the "Grey" palette displays various shades of grey, ranging from dark to light.},width=.7\textwidth,keepaspectratio]{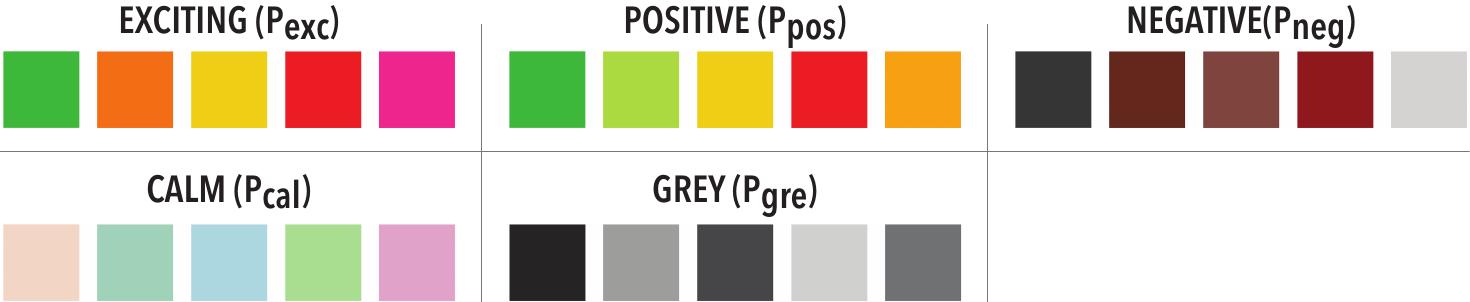}
  \captionof{figure}{Color palettes used in S1.}
  \label{fig:s1_graphs}
  \vspace{1.5em}
  \includegraphics[alt={This figure shows the mean arousal and valence ratings for each color palette in Study 1, centered on a neutral value of 5. For arousal, the exciting palette (P_exc) has a mean of 3.81 [3.20, 4.40], the positive palette (P_pos) is 6.88 [6.53, 7.25], the negative palette (P_neg) is 2.86 [2.38, 3.31], the calm palette (P_calm) is 3.80 [3.28, 4.31], and the grey palette (P_grey) is 5.53 [4.98, 6.09]. For valence, the exciting palette has a mean of 6.63 [6.16, 7.12], the positive palette is 6.65 [6.45, 7.29], the negative palette is 2.98 [2.73, 3.25], the calm palette is 4.71 [4.19, 5.19], and the grey palette is 6.49 [5.96, 7.03]. Error bars indicate the 95\% confidence interval for each condition.},width=.85\textwidth,keepaspectratio]{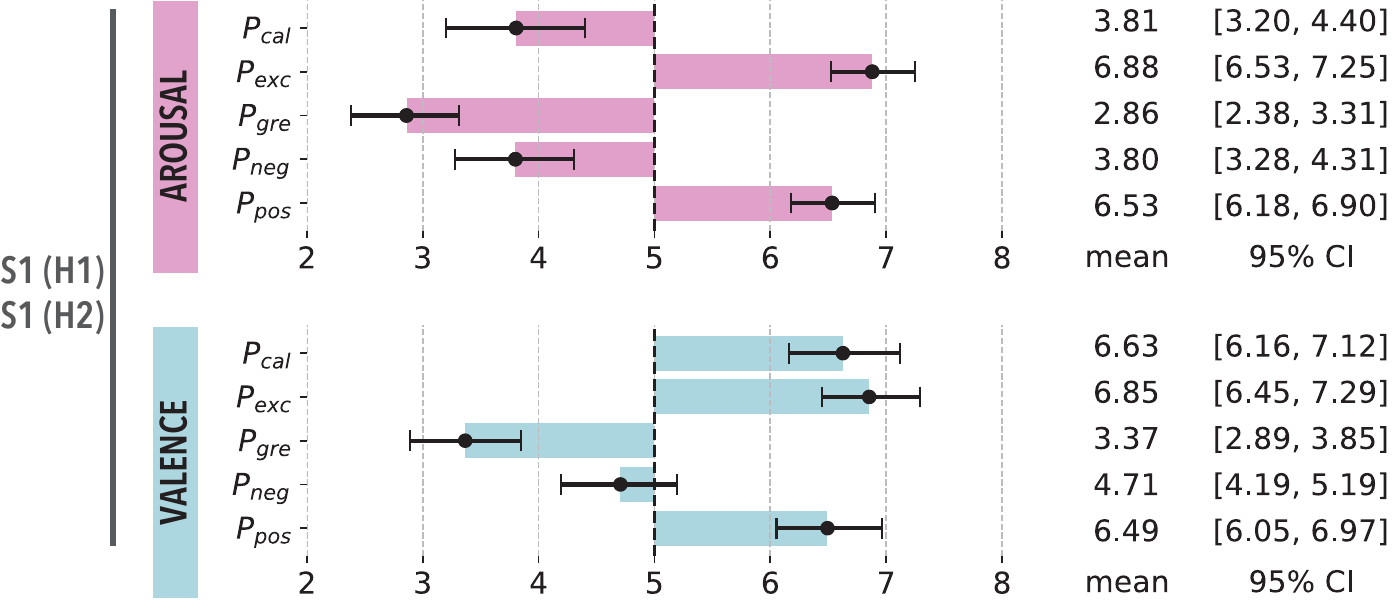}
  \captionof{figure}{Mean results for each color palette for S1, centered on a neutral value of $5$. \errorbardescrip}
  \label{fig:s1_res_all}
  \vspace{0.75em}
   \includegraphics[alt={This figure displays a set of charts representing nine conditions in Study 2, combining three data trends (negative, neutral, and positive) with three color palettes (exciting, grey, and calm). The charts are arranged in a grid format where each row corresponds to a color palette, and each column corresponds to a data trend. The exciting palette uses vibrant colors, the grey palette uses various shades of grey, and the calm palette uses pastel colors. The direction and pattern of the data points within each chart visually represents the data trends.}, width=.65\textwidth,keepaspectratio]{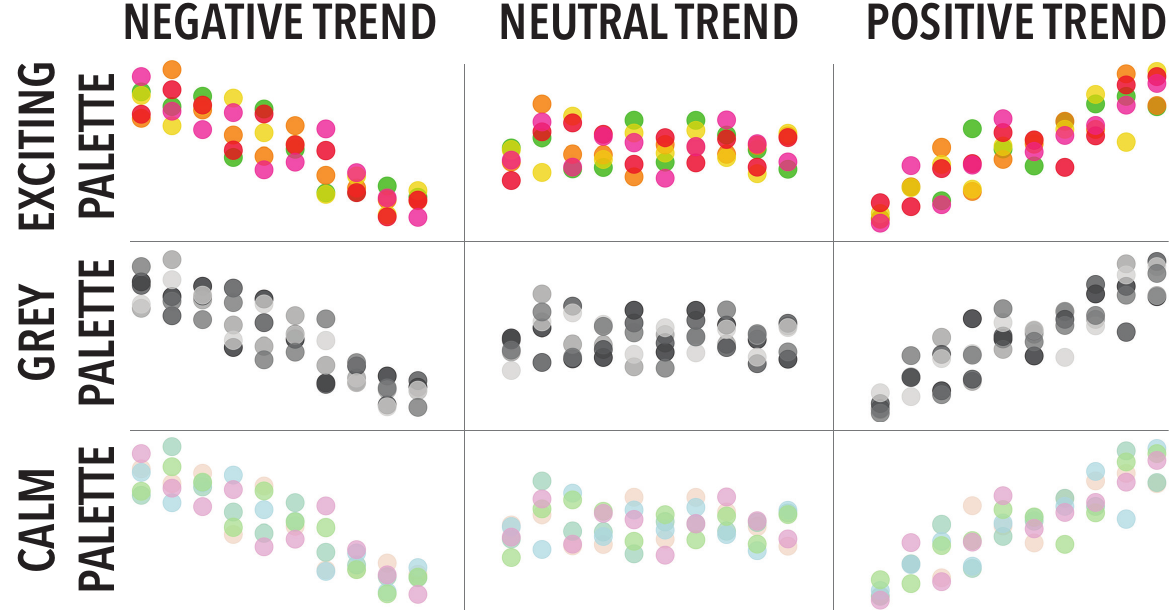}
  \captionof{figure}{One set of charts for all 9 conditions used in S2 (three data trends and three color palettes).}
  \label{fig:s2_graphs}
   \end{minipage}
   \vspace{-1.0em} 
\end{figure}

\inlinesection{Datasets}
Like the four color palettes from previous work~\cite{Bartram:2017:ACV}, 
our grey-scale palette has five values, set by linear interpolation between $0$ and $1$ on the lightness dimension of the CIELAB color space: $(0.166,0,0), (0.333, 0, 0), (0.5, 0, 0), (0.666, 0, 0),$ and $(0.833, 0, 0)$. 
\Cref{fig:s1_graphs} shows the color palettes used in this study.

\inlinesection{Participants}
50 participants participated. 
They took on average 8 minutes 9 seconds to complete the study and were compensated \pounds 1.5. 

\subsection{S1 -- Hypotheses}
We formulated two hypotheses for S1:
\begin{description}[noitemsep]
    \item[H1:] $P_{cal}$, $P_{exc}$, $P_{neg}$ and $P_{pos}$ will induce the same emotion as in Bartram \etal's work~\cite{Bartram:2017:ACV}: 
    \arousalCol{$P_{cal}$} will be low and \valenceCol{$P_{cal}$} high; 
    \arousalCol{$P_{exc}$} and \valenceCol{$P_{exc}$} high; 
    \arousalCol{$P_{neg}$} and \valenceCol{$P_{neg}$} low; and 
    \arousalCol{$P_{pos}$} and \valenceCol{$P_{pos}$} high.

    \item[H2:] \arousalCol{$P_{gre}$} and \valenceCol{$P_{gre}$} will have the most neutral values of the five palettes because the other palettes were designed to induce emotion via color and $P_{gre}$ can be seen as having no color.
\end{description}

\subsection{S1 -- Results}
No responses were excluded. In total, we gathered 1500 trial answers (50 participants $\times$ 5 color palettes  $\times$ 2 measurements $\times$ 3 repetitions). 

To answer \textbf{H1} and \textbf{H2} we look at \Cref{fig:s1_res_all}. 
\textbf{H1} is confirmed, with 
\arousalCol{$P_{cal}$} being low and \valenceCol{$P_{cal}$} being high; 
both \arousalCol{$P_{exc}$} and \valenceCol{$P_{exc}$} being high; 
\arousalCol{$P_{neg}$} being low and \valenceCol{$P_{neg}$} being neutral to low; and
both \arousalCol{$P_{pos}$} and \valenceCol{$P_{pos}$} being high.
\textbf{H2} is rejected, with both \arousalCol{$P_{gre}$} and \valenceCol{$P_{gre}$} being low.

\subsection{S1 -- Discussion}
Confirming \textbf{H1} (our results are consistent with those from Bartram \etal~\cite{Bartram:2017:ACV}) tells us that we can use the SAM scales to capture emotions and that our results can be used as benchmarks for future studies.

Although we did not establish a neutral palette (\textbf{H2} rejected),
$P_{gre}$ is an appropriate baseline for studying the amount of emotion induced by other factors, given it has the lowest effect on emotion overall.


\section{S2 -- Color Palettes and Trends}
\label{sec:S2}
S1 showed color palettes outside a data visualization context, but data visualizations that show real datasets exhibit some characteristics. 
One such characteristic is the data trend, which can give the impression of motion, which we know affects emotions (\eg~\cite{Bartram:2010:WMM, Lockyer:2012:AMT}).
In this study, we look at the effect of both color and trend on emotions.

\subsection{S2 -- Method}
\label{sec:s2_method}

\inlinesection{Design}
We used the most extreme palette in each quadrant because they are the most representative of a certain emotion: the 
exciting (\arousalCol{$P_{exc}$} high, \valenceCol{$P_{exc}$} high), 
calm (\arousalCol{$P_{cal}$} low, \valenceCol{$P_{cal}$} high), and 
grey (\arousalCol{$P_{gre}$} low, \valenceCol{$P_{gre}$} low) palettes.
We only used three palettes because S1 did not reveal any color palette having high arousal and low valence.
We combined them with three trends: positive ($T_{pos}$), neutral ($T_{neu}$), and negative ($T_{neg}$). 
Three color palettes and three trends resulted in nine conditions.
With two repeated measures for each condition, there were two arousal blocks and two valence blocks.
Each block had nine graphs -- one for each of the nine conditions. 
We used a different dataset for each repetition, generated as explained in the next section.
We used scatterplots to represent the datasets because they are commonly used to 
represent data trend, with a balanced sampling of each color from a given palette.

\inlinesection{Datasets}
We generated the experimental datasets with negative, neutral, and positive trends using the following procedure:
\begin{enumerate}[noitemsep]
    \item Define a desired correlation, $\rho$, and define $\theta = \arccos(\rho)$;
    \item Define a vector $x_1$ for $x$-coordinates and a random vector $x_2$ sampled from $\mathcal{N}(2, 0.5)$ of the same length;
    \item Center $x_1$ and $x_2$ so that they have mean $0$, yielding $\dot{x}_1$ and $\dot{x}_2$;
    \item Make $\dot{x}_2$ orthogonal to $\dot{x}_1$ by projecting it onto the orthogonal subspace defined by $\dot{x}_1$, yielding $\dot{x}_2^{\perp}$;
    \item Scale $\dot{x}_1$ and $\dot{x}_2^{\perp}$ to length $1$, yielding $\bar{x}_1$ and $\bar{x}_2^{\perp}$;
    \item Lastly define a vector $y = \bar{x}_2^{\perp} + \frac{1}{\tan{\theta}}\cdot\bar{x}_1$;
    \item $x_1$ and $y$ now have the desired correlation $\rho$.
\end{enumerate}

We set $\rho$ values of $-0.9$ for $T_{neg}$, $0$ for $T_{neu}$, and $0.9$ for $T_{pos}$. 
We did not choose $\rho$ values of $-1$ and $1$ 
so that the plots exhibit some realism.
To each $x$-coordinate corresponds five data points -- one for each color of the palette.
\Cref{fig:s2_graphs} shows one set of graphs used in this study.

\inlinesection{Procedure}
In addition to the procedure described in \Cref{sec:method}, at the end of each block, participants rated whether their responses were affected more by the colors or the trend of the graph, on a 5-point Likert-scale ranging from ``mainly color'' to ``mainly trend''.

\inlinesection{Participants}
50 participants participated. 
They took on average 7 minutes 11 seconds to complete the study and were compensated \pounds 1.2.

\subsection{S2 -- Hypotheses}
We formulated four hypotheses for S2:
\begin{description}[noitemsep]
    \item[H1:] arousal and valence for each color palette will be consistent with results from S1, as the data has no meaning and we have controlled for trend by averaging over negative, neutral, and positive trends.

    \item[H2:] \valenceCol{$T_{pos} > T_{neu}$} and \valenceCol{$T_{neu} > T_{neg}$} because positive trends generally represent good things while negative trends generally represent bad things. Further, Bartram and Nakatani~\cite{Bartram:2010:WMM} found that abstract motions in the upper-side of a given space are rated positively.

    \item[H3:] arousal will not be affected by trend because we did not gather indications from previous research of such an effect.

    \item[H4:] arousal and valence will be more affected by color (i.e. will lead to more variance) than by trend, as there is a well-established, clear effect of color on emotion.

\end{description}

\begin{figure}[!t]
\begin{minipage}[t]{\columnwidth}
 \centering

  \includegraphics[alt={This figure illustrates the differences in arousal and valence ratings between Study 1 (S1) and Study 2 (S2) for different color palettes, with means and 95\% confidence intervals. For the exciting palette (P_exc), the arousal difference is 0.88 [0.18, 1.57] and the valence difference is 0.43 [-0.11, 0.96]. The calm palette (P_calm) shows an arousal difference of 0.08 [-0.51, 0.64] and a valence difference of 1.17 [0.62, 1.75]. The grey palette (P_grey) exhibits an arousal difference of 0.13 [-0.39, 0.66] and a valence difference of -0.40 [-0.95, 0.16].},width=.88\textwidth,keepaspectratio]{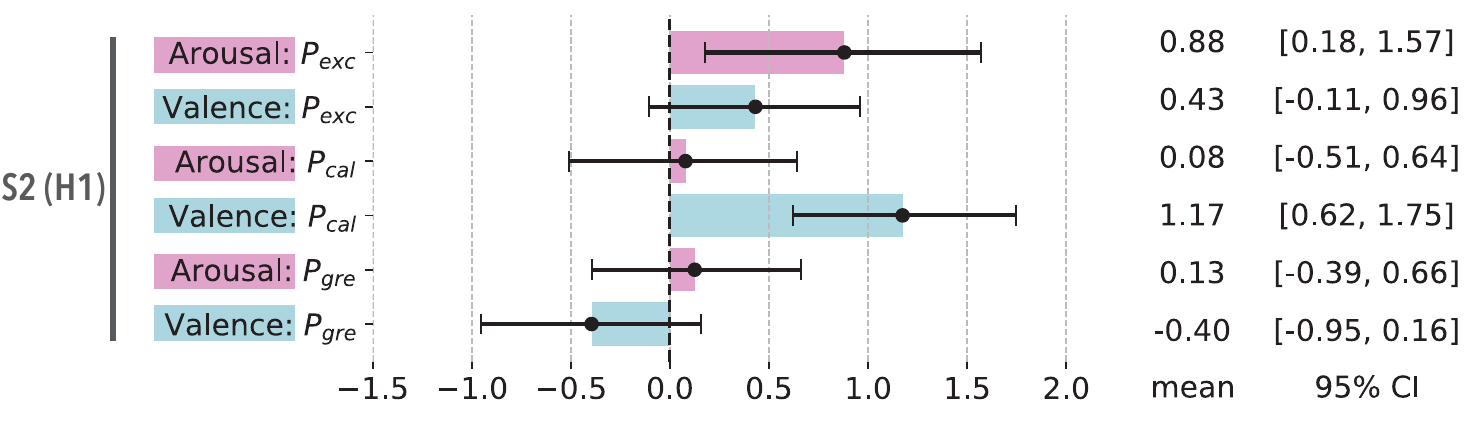}
  \captionof{figure}{Pairwise differences between results from S1 and S2. \errorbardescrip}
  \label{fig:s2_res_color_pair}

  \vspace{1.5em}

  \includegraphics[alt={This figure presents the mean arousal and valence ratings for different color palettes from Study 2 (opaque colors) and Study 1 (light grey). For the exciting palette, S2 shows a higher arousal (mean around 7) and similar valence (mean around 7) compared to S1. The calm palette in S2 shows a lower arousal (mean around 4.5) and slightly higher valence (mean around 6) than in S1. The grey palette shows consistent arousal and valence ratings between S2 (mean arousal around 3.5, valence around 4) and S1. Error bars represent the 95\% confidence intervals for each condition.}, width=.8\textwidth,keepaspectratio]{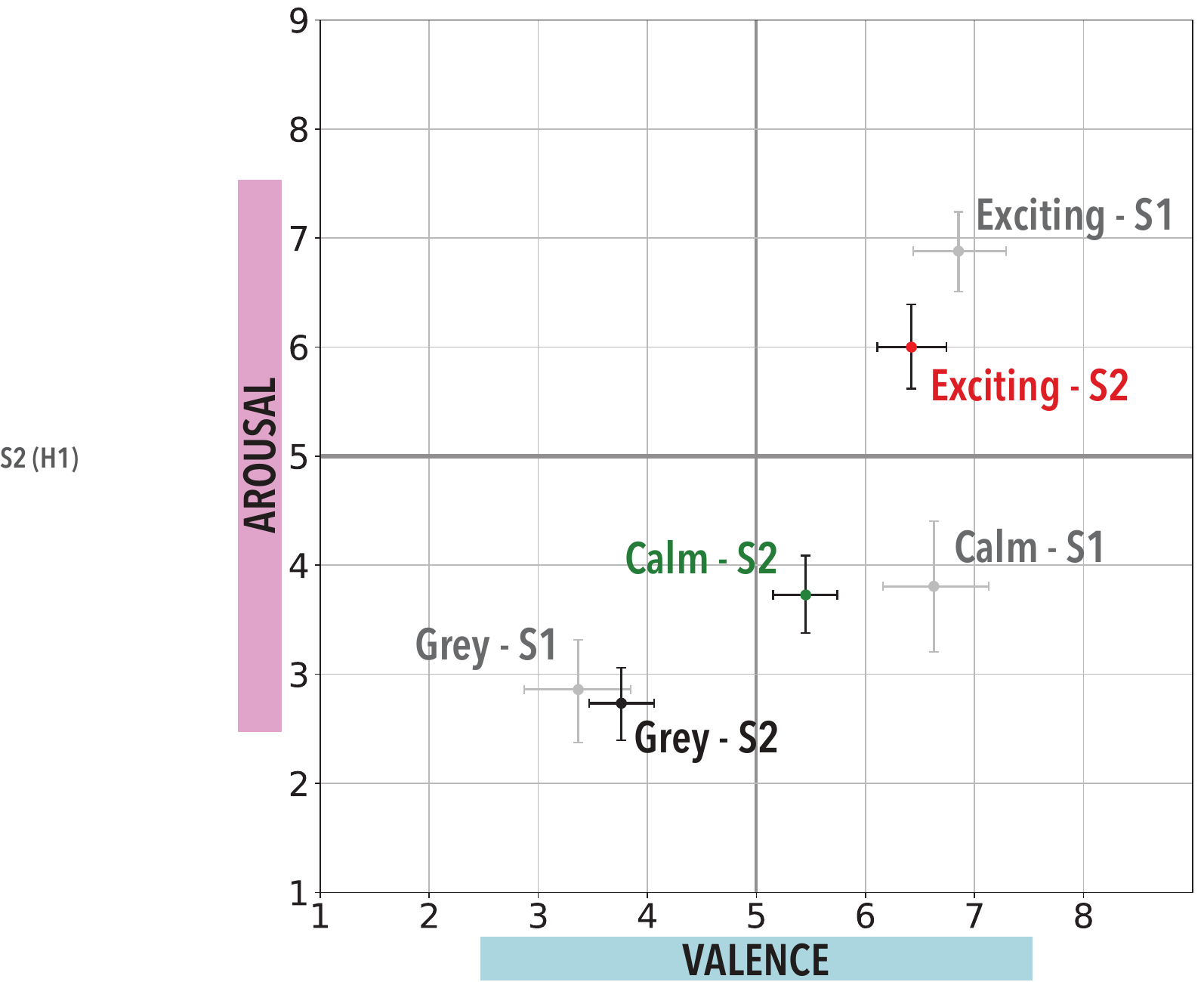}
  \captionof{figure}{Mean arousal and valence for color palettes from S2 (opaque color, suffixed with -S2) and S1 (light grey, suffixed with -S1). \errorbardescrip}
  \label{fig:s2_res_color_mean}
  
  \vspace{1.5em}

  \includegraphics[alt={This figure illustrates the differences in arousal and valence ratings between various trends for all color palettes in Study 2. The arousal differences are shown in pink bars, where T_pos - T_neu has a mean difference of 0.48 [0.32, 0.64] and T_neu - T_neg shows a mean difference of 0.18 [0.01, 0.33]. The valence differences, shown in blue bars, indicate that T_pos - T_neu has a mean difference of 0.39 [0.23, 0.55] and T_neu - T_neg has a mean difference of 0.22 [0.06, 0.38]. Error bars represent the 95\% confidence intervals for each condition.},width=.88\textwidth,keepaspectratio]{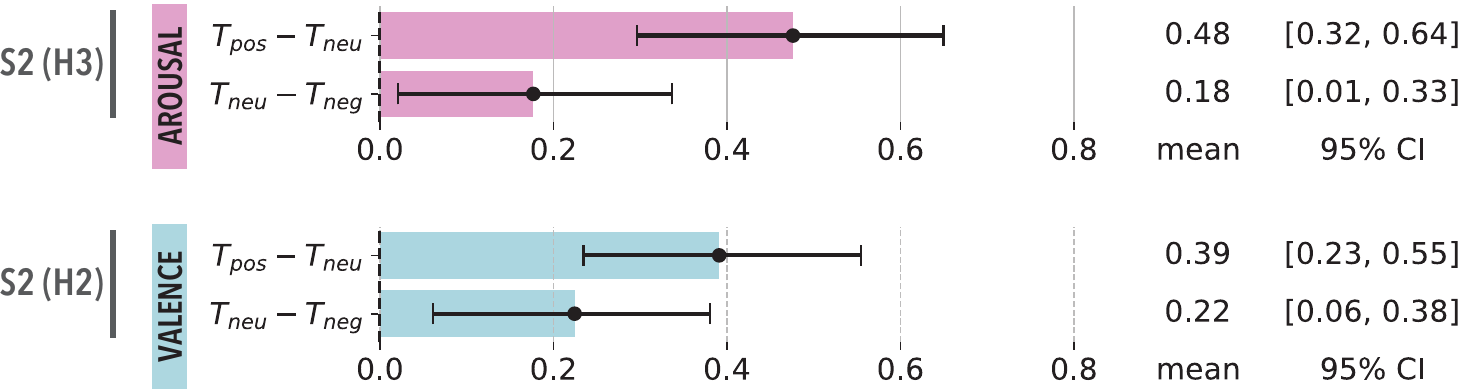}
  \captionof{figure}{S2: Differences between trends (all color palettes). \errorbardescrip}
  \label{fig:s2_res_trends}
  
  \vspace{1.5em}

  \includegraphics[alt={This figure shows the pairwise differences in standard deviations (σ) for color and trend in arousal and valence ratings in Study 2. The arousal difference is indicated by a pink bar with a mean of 1.37 and a 95\% confidence interval of [1.14, 1.61]. The valence difference is represented by a blue bar with a mean of 1.11 and a 95\% confidence interval of [0.89, 1.33]. Error bars represent the 95\% confidence intervals for each measurement.},width=.88\textwidth,keepaspectratio]{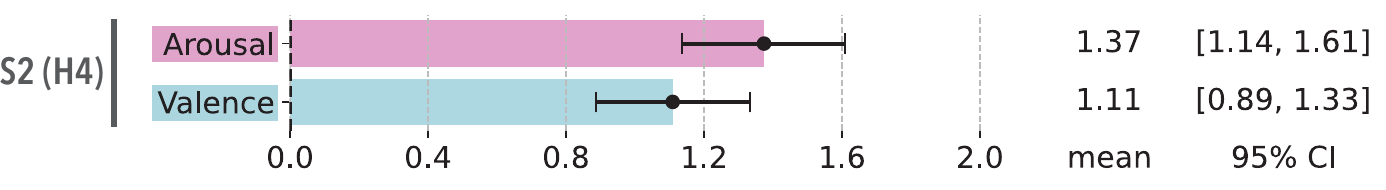}
  \captionof{figure}{S2: Pairwise difference between $\sigma_{Color}$ and $\sigma_{Trend}$. \errorbardescrip}
  \label{fig:s2_res_diff_color_trend}
    
  \vspace{1.5em}

  \includegraphics[alt={This figure illustrates participants' self-reported ratings on whether trends or color affect their choices more, measured on a scale from 1 (mainly color) to 5 (mainly trend). For arousal, 30\% rated at 1, 12\% at 2, 22\% at 3, 12\% at 3.5, and 4\% at 4 and 4.5. For valence, 24\% rated at 1, 16\% at 2, 20\% at 3, 6\% at 3.5, 16\% at 4, and 8\% at 5.},width=1.0\textwidth,keepaspectratio]{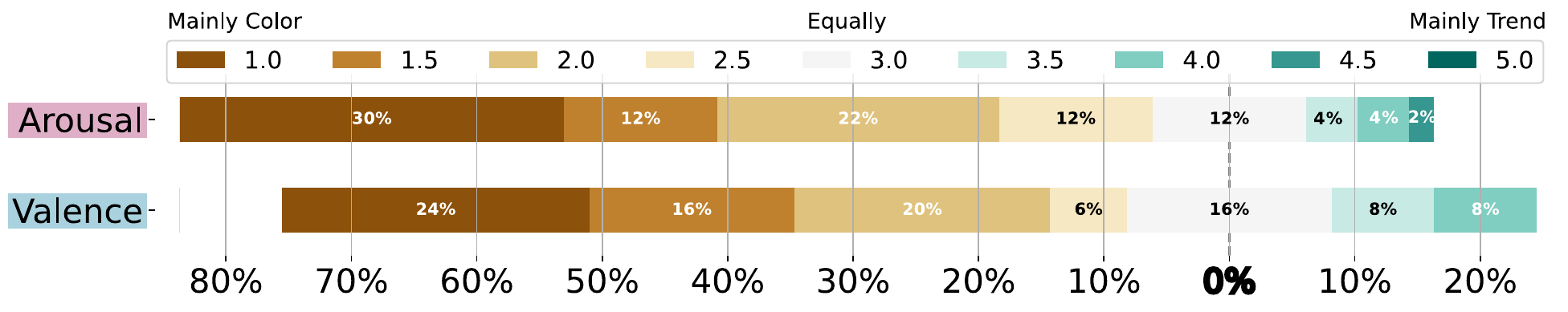}
  \captionof{figure}{S2: Participants' self-reported ratings on whether trends or color affect their choices more. Percentages are rounded.}
  \label{fig:s2_res_likert}

 \end{minipage}
\vspace{-1em} 
\end{figure}

\subsection{S2 -- Results}\label{sec:s2:results}
We discarded the data from one participant who provided the same answers for all trials.
In total, we gathered 1750 trial answers (49 participants $\times$ 3 color palettes $\times$ 3 trends $\times$ 2 measurements $\times$ 2 repetitions) and 196 Likert-scale ratings (49 participants $\times$ 4 blocks).

\inlinesection{Effect of color}
To answer \textbf{H1}, we examine pairwise differences between the ratings of palettes from S1 and from S2, regardless of trend (see \Cref{fig:s2_res_color_pair}).
The CI for \arousalCol{$P_{exc}$} does not overlap with 0, which is strong evidence that \arousalCol{$P_{exc}$} is greater in S1 than in S2. On the other hand, the CI for \valenceCol{$P_{exc}$} slightly overlaps with 0, so there is no evidence of such difference for \valenceCol{$P_{exc}$}.
\arousalCol{$P_{cal}$} is similar in both studies, but there is strong evidence that \valenceCol{$P_{cal}$} is greater in S1 than in S2.
Both \arousalCol{$P_{gre}$} and \valenceCol{$P_{gre}$} are similar in both studies.
\Cref{fig:s2_res_color_mean} summarizes these results.

\inlinesection{Effect of trend}
To answer \textbf{H2} and \textbf{H3}, for each measure, we compute and bootstrap the signed difference between each pair of trends (see \Cref{fig:s2_res_trends}).
There is strong evidence that \arousalCol{$T_{pos} > T_{neu}$} but the effect is small given the measurable range from $-8$ to $+8$. This effect is even smaller for \arousalCol{$T_{neu} > T_{neg}$}.
There is similar evidence that \valenceCol{$T_{pos} > T_{neu}$} and \valenceCol{$T_{neu} > T_{neg}$}.
By transitivity, \arousalCol{$T_{pos} > T_{neu} > T_{neg}$} and \valenceCol{$T_{pos} > T_{neu} > T_{neg}$}.

\inlinesection{Effect of color and trend}
To answer \textbf{H4}, we analyze the differences between the standard deviations $\sigma$ for color and trend (the larger $\sigma$, the larger the effect). 
We first computed
\arousalCol{$\sigma_{color} = \{G_{exc}, G_{cal}, G_{gre}\}$},
\valenceCol{$\sigma_{color} = \{G_{exc}, G_{cal}, G_{gre}\}$},
\arousalCol{$\sigma_{trend} = \{T_{pos}, T_{neu}, T_{neg}\}$}, and
\valenceCol{$\sigma_{trend} = \{T_{pos}, T_{neu}, T_{neg}\}$} for each participant.  
Then, we bootstrapped the differences \arousalCol{$\sigma_{color} - \sigma_{trend}$} and \valenceCol{$\sigma_{color} - \sigma_{trend}$} for each participant (see \Cref{fig:s2_res_diff_color_trend}).
There is strong evidence that \arousalCol{$\sigma_{color} > \sigma_{trend}$} and \valenceCol{$\sigma_{color} > \sigma_{trend}$}, and the differences are quite large.

\inlinesection{Self-reported perceptions}
\Cref{fig:s2_res_likert} shows that over two-thirds of the participants found that color affected both measures more than trend. 

\subsection{S2 -- Discussion}
\textbf{H1} is only partially confirmed because we found differences between the results from S1 and S2, namely for \arousalCol{$P_{exc}$} and for \valenceCol{$P_{cal}$}. 
However, as all color palettes remain in their same quadrant in the arousal-valence chart (see \Cref{fig:s2_res_color_mean}), the conveyed emotion of each palette remains consistent. 
We thus conclude that although there are small variations with and without data, color palettes can be studied without data visualizations and still provide meaningful results. 
This also makes an important link between the study of color in psychology and in visualization, as the presentation of color stimuli in the two fields generally differs~\cite{Bartram:2017:ACV, wilms2018color}.

Regarding the effect of trend in isolation, the results support \textbf{H2} but not \textbf{H3}. 
Yet, we interpret this result with caution because the effect sizes are relatively small compared to the effects of color. 

The results strongly support \textbf{H4} for both measurements. 
Our discussion of \textbf{H2} and \textbf{H3} as well as participants' self-reported perceptions demonstrate that color has a greater impact on emotions than trend. 

In addition to our confirmatory analysis, through exploratory analysis, we found \arousalCol{$T_{pos} - T_{neu} > T_{neu} - T_{neg}$} and \valenceCol{$T_{pos} - T_{neu} > T_{neu} - T_{neg}$} for all the palettes we tested. 
This suggests that the effect of trend on emotion is not linear and that this effect should be particularly considered in visualizations that exhibit strong positive data trends.


\section{S3 -- Chart Types and Data Trends}
\label{sec:S3}
We conducted S2 with scatterplots for their common usage and ability to represent trends. 
As we know, chart types influence people's perception of shapes and trends in data, which in turn affects people's emotions. In S3, we study the effects of chart type and data trend on emotions. 

\subsection{S3 -- Method}

\inlinesection{Design}
We determined which charts to study so that they would: 
1) be widely used in visualization;
2) be well-suited to conveying data trends; and
3) support adjusting the amount of ink that is being used to represent data, so that we can mitigate the potential confound that color can create, by using the same amount of ink for all charts.
Using these criteria, we selected four types of charts: bar chart ($C_{bar}$), scatterplot ($C_{sca}$), jagged line chart ($C_{lin}$), and smoothed line chart ($C_{sml}$). 
We included two types of line charts because this allows us to study in a visualization context the well-known differences in emotional response to jagged vs smooth objects~\cite{aryani2020affective,bar2006humans}.
We kept the three trends used in S2: $T_{pos}$, $T_{neu}$, and $T_{neg}$. 
Four types of charts and three trends resulted in twelve conditions.
With three repeated measures for each condition, there were three arousal blocks and three valence blocks. 
Each block had twelve graphs -- one for each of the twelve conditions.
We used a different dataset for each repetition, generated as explained next.

\begin{figure}[!t]
\begin{minipage}[!t]{\columnwidth}
 \centering

    \includegraphics[alt={This figure presents 12 conditions combining three data trends (negative, neutral, positive) and four chart types (bar chart, jagged line chart, scatterplot, smoothed line chart) from Study 3. Each row represents a different chart type, and each column represents a different data trend. The x-axis for all charts is labeled "time."},width=.687\columnwidth,keepaspectratio]{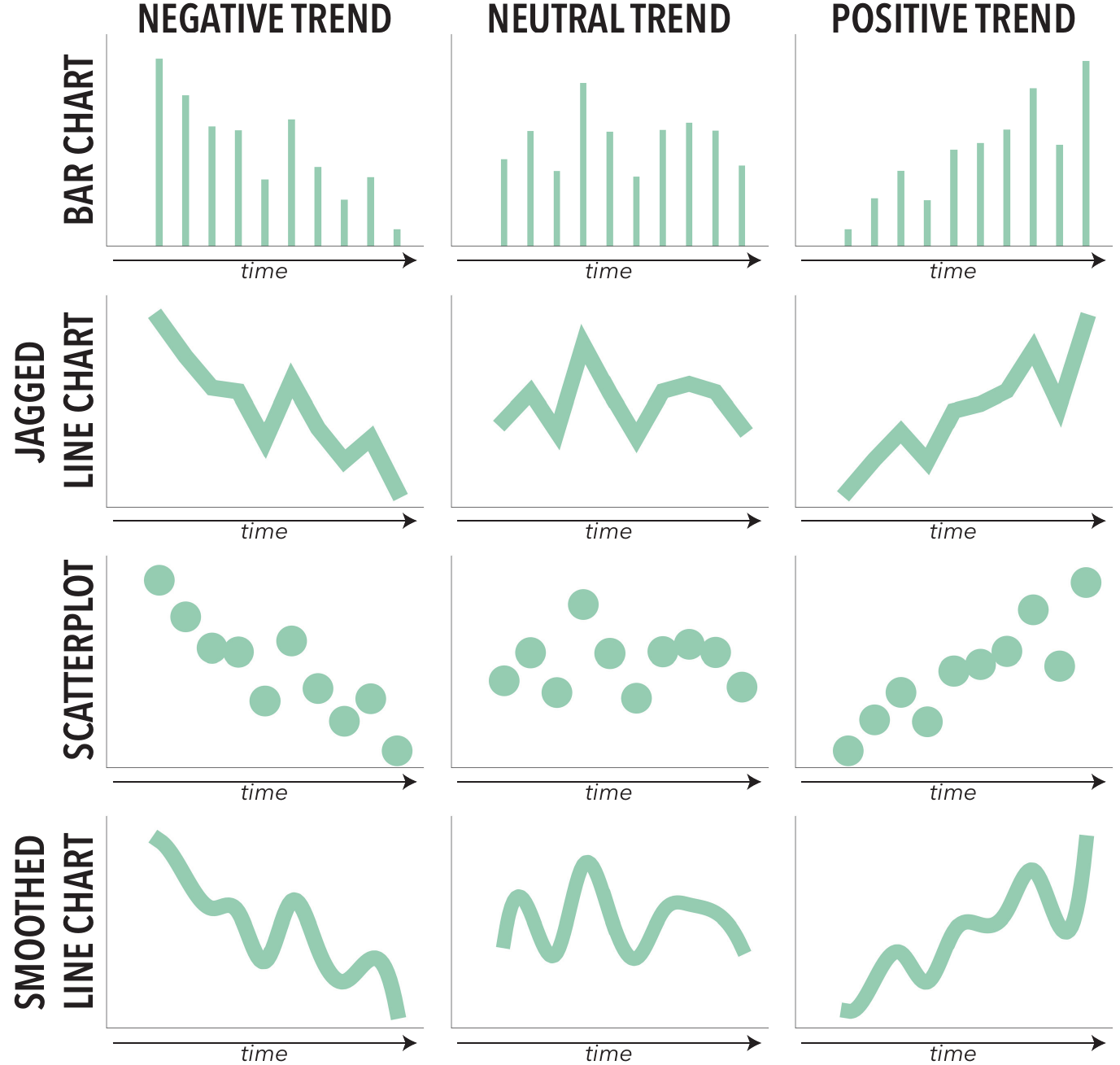}
  \captionof{figure}{One set of charts for all 12 conditions used in S3 (three data trends and four chart types), with a ``Time'' label on the $y$ axis.}
  \label{fig:s3_graphs}

   \vspace{1.5em}

  \includegraphics[alt = {This figure presents the differences in arousal and valence ratings between various chart types across all trends in Study 3. For arousal (pink bars), comparisons show the mean differences and 95 percent confidence intervals: bar charts versus line charts (-0.68 [-0.98, -0.32]), bar versus smoothed line (-0.70 [-1.08, -0.24]), bar versus scatterplot (-0.40 [-0.74, -0.01]), line versus smoothed line (-0.02 [-0.28, 0.35]), line versus scatterplot (0.28 [-0.00, 0.60]), and smoothed line versus scatterplot (0.30 [0.10, 0.50]). For valence (blue bars), the differences are: bar versus line (0.10 [-0.14, 0.31]), bar versus smoothed line (-0.22 [-0.56, 0.17]), bar versus scatterplot (-0.11 [-0.50, 0.28]), line versus smoothed line (-0.32 [-0.60, 0.03]), line versus scatterplot (-0.12 [-0.53, 0.13]), and smoothed line versus scatterplot (0.12 [-0.13, 0.33]). Error bars represent the 95 percent confidence intervals for each condition.},width=.88\textwidth,keepaspectratio]{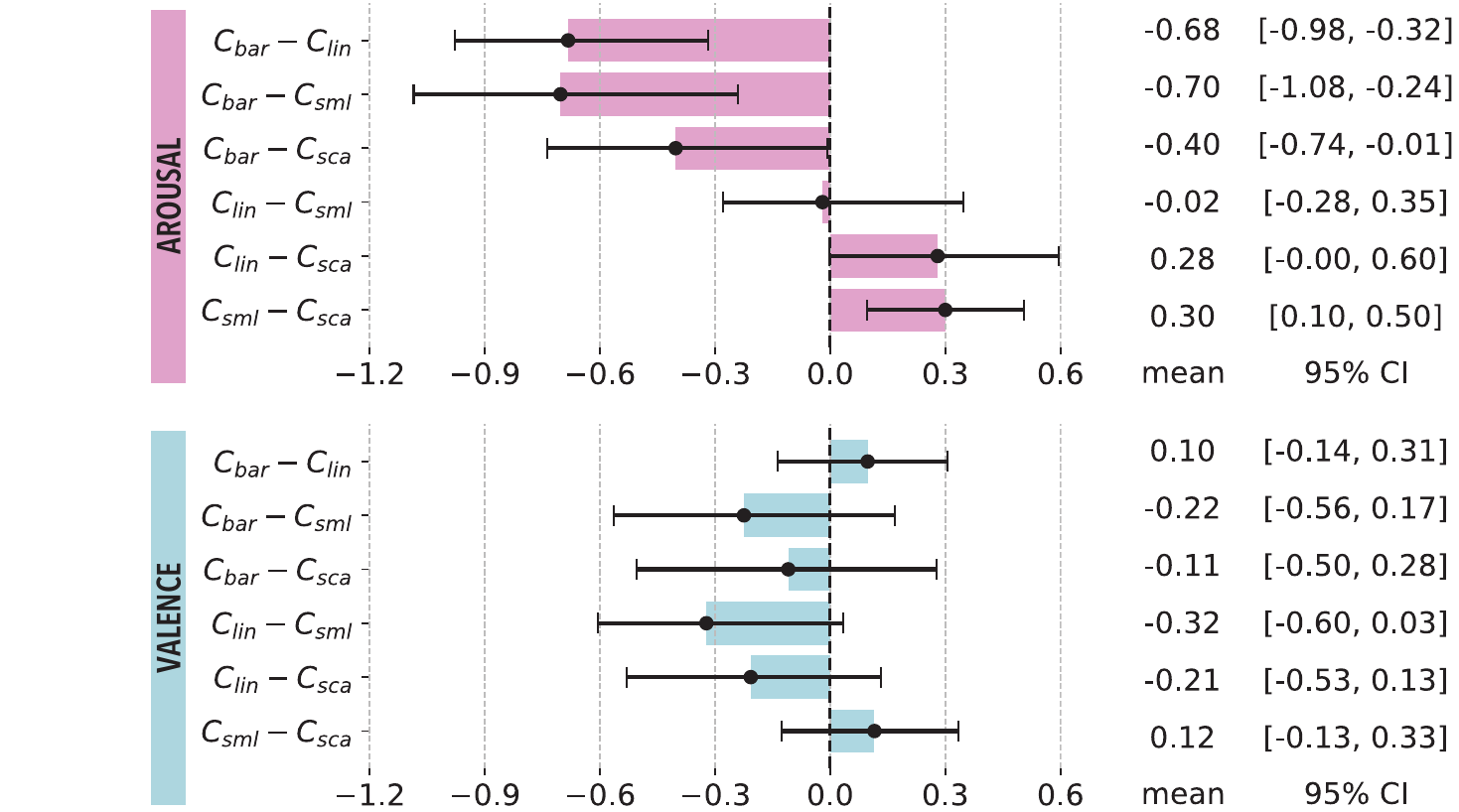}
  \captionof{figure}{S3: Differences between chart types (all trends). \errorbardescrip}
  \label{fig:s3_res_charts}
  
  \vspace{1.5em}

  \includegraphics[alt = {This figure illustrates the differences in arousal and valence ratings between various trends across all chart types in Study 3. For arousal (pink bars), the mean differences and 95 percent confidence intervals are: T_pos - T_neu (1.48 [1.14, 1.82]) and T_neu - T_neg (0.56 [0.09, 1.02]). For valence (blue bars), the differences are: T_pos - T_neu (1.53 [1.18, 1.86]) and T_neu - T_neg (1.23 [0.76, 1.68]). Error bars represent the 95 percent confidence intervals for each condition.}, width=.88\textwidth,keepaspectratio]{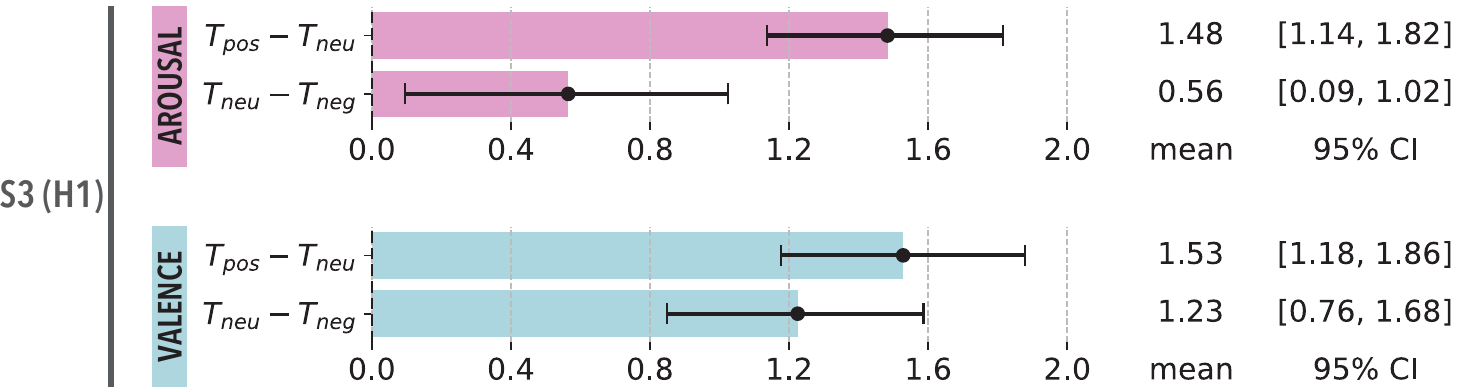}
  \captionof{figure}{S3: Differences between trends (all chart types). \errorbardescrip}
  \label{fig:s3_res_trends}

  \vspace{1.5em}

  \includegraphics[alt={This figure presents the differences in arousal and valence ratings for T_pos - T_neu and T_neu - T_neg across different chart types in Study 3.For arousal in T_pos - T_neu (pink bars), the mean differences and 95 percent confidence intervals are: bar versus line (0.20 [-0.13, 0.52]), bar versus smoothed line (-0.21 [-0.48, 0.03]), bar versus scatterplot (0.01 [-0.37, 0.40]), line versus smoothed line (-0.19 [-0.41, 0.05]), line versus scatterplot (0.03 [-0.27, 0.33]), and smoothed line versus scatterplot (0.03 [-0.27, 0.33]). For valence in T_pos - T_neu (blue bars), the differences are: bar versus line (-0.43 [-0.64, -0.22]), bar versus smoothed line (-0.37 [-0.63, -0.13]), bar versus scatterplot (-0.33 [-0.55, -0.10]), line versus smoothed line (0.05 [-0.15, 0.25]), line versus scatterplot (0.10 [-0.11, 0.30]), and smoothed line versus scatterplot (0.05 [-0.19, 0.29]). For arousal in T_neu - T_neg (pink bars), the differences are: bar versus line (-0.21 [-0.51, 0.08]), bar versus smoothed line (-0.21 [-0.51, 0.11]), bar versus scatterplot (-0.08 [-0.44, 0.27]), line versus smoothed line (0.01 [-0.24, 0.26]), line versus scatterplot (0.13 [-0.13, 0.38]), and smoothed line versus scatterplot (0.13 [-0.16, 0.41]). For valence in T_neu - T_neg (blue bars), the differences are: bar versus line (-0.12 [-0.38, 0.13]), bar versus smoothed line (-0.01 [-0.37, 0.39]), bar versus scatterplot (-0.13 [-0.25, 0.29]), line versus smoothed line (0.11 [-0.18, 0.43]), line versus scatterplot (0.15 [-0.13, 0.43]), and smoothed line versus scatterplot (0.03 [-0.38, 0.41]). Error bars represent the 95 percent confidence intervals for each condition.},width=.88\textwidth,keepaspectratio]{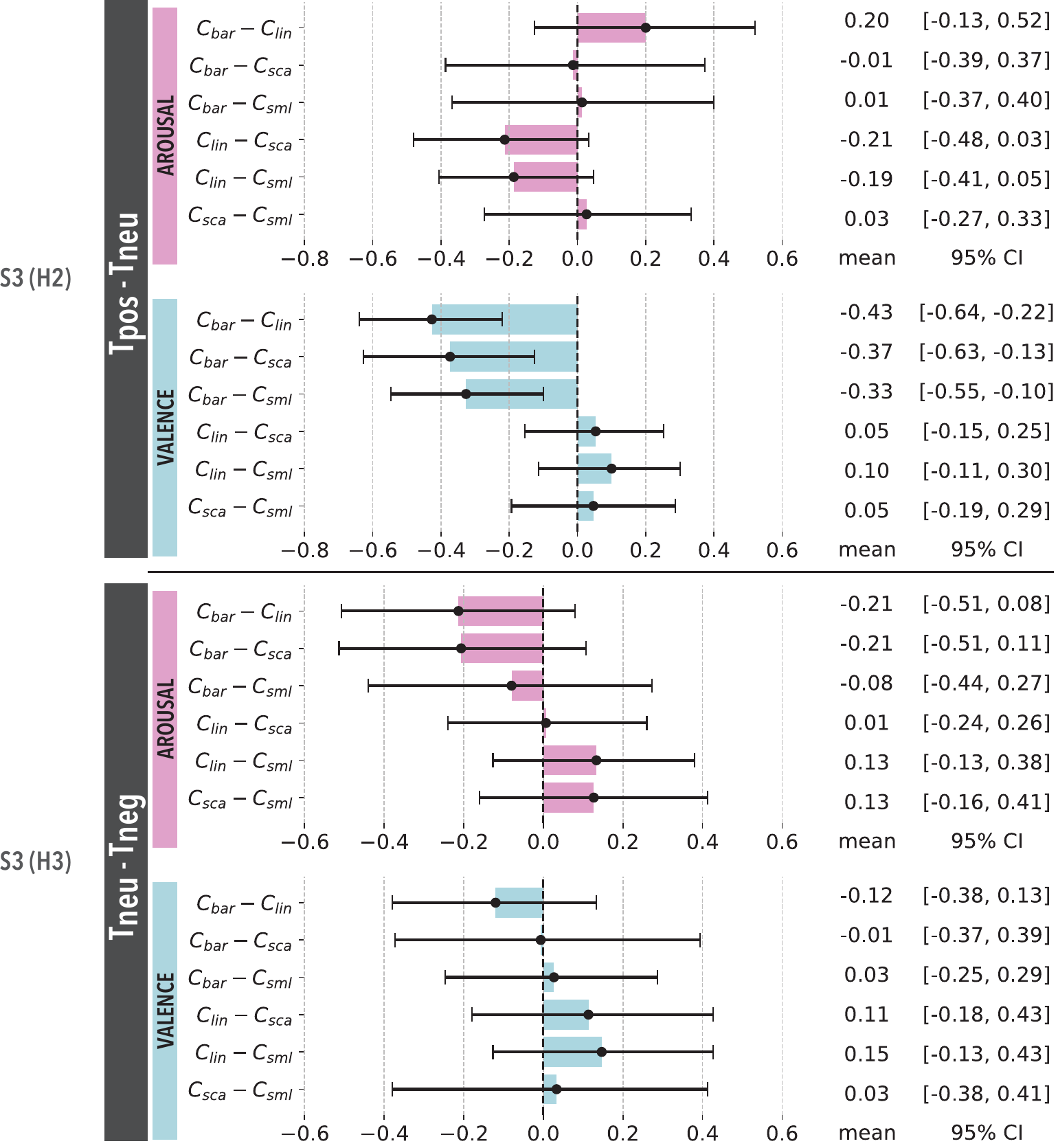}
  \captionof{figure}{S3: $T_{pos} - T_{neu}$ and $T_{neu} - T_{neg}$ differences between chart types. \errorbardescrip}
  \label{fig:s3_res_posneu_neuneg}
   
  \vspace{-3.0em}
 \end{minipage}
\end{figure}

\inlinesection{Datasets}
We generated the datasets with different trends as explained in \Cref{sec:s2_method}. 
To control the amount of ink used in each chart, we varied the visual elements of the charts, calculated the percent of colored pixels for each variation, and selected the one with the desired percentage (6\% in this study).
\Cref{fig:s3_graphs} shows one set of 12 charts.

\inlinesection{Procedure}
In addition to the procedure described in \Cref{sec:method}, at the end of the study, participants rated whether their responses were affected more by chart type or by the trend shown in the chart, on a 5-point Likert-scale ranging from ``mainly type'' to ``mainly trend''. 

\inlinesection{Participants}
50 participants participated. 
They took on average 11 minutes 50 seconds to complete the study and were compensated \pounds 1.65. 

\subsection{S3 -- Hypotheses}
We formulated six hypotheses for S3:
\begin{description}[noitemsep]
    \item[H1:] when aggregating over all chart types, \arousalCol{$T_{pos}>T_{neu}>T_{neg}$} and \valenceCol{$T_{pos}>T_{neu}>T_{neg}$}, following results from S2.
    
    \item[H2:] \arousalCol{$T_{pos} - T_{neu}$} and \valenceCol{$T_{pos} - T_{neu}$} will be greater for $C_{lin}$, $C_{sml}$ and $C_{sca}$ than for $C_{bar}$ because they make trends more noticeable~\cite{10.1145/3025453.3025922}.
    
    \item[H3:] \arousalCol{$T_{neu} - T_{neg}$} and \valenceCol{$T_{neu} - T_{neg}$} will be greater for $C_{lin}$, $C_{sml}$ and $C_{sca}$ than for $C_{bar}$, for the same reason.

    \item[H4:] \arousalCol{$C_{lin} > C_{sml}$}, as jagged shapes are associated with higher arousal than smooth shapes~\cite{aryani2020affective,bar2006humans}.

    \item[H5:] \valenceCol{$C_{lin} < C_{sml}$}, as jagged shapes are associated with lower valence than smooth shapes~\cite{aryani2020affective,bar2006humans}.
    
    \item[H6:] arousal and valence will be more affected by trend than by chart type, as results from S2 show that the trend affects emotion, while there is no known evidence of effects of chart types on emotions.
\end{description}

\subsection{S3 -- Results}
No responses were excluded. 
In total, we gathered 3600 trial answers (50 participants × 4 chart types × 3 trends × 2 measurements × 3 repetitions) and 100 Likert-scale ratings (50 participants × 2 blocks). 

\inlinesection{Effect of chart types}
We look at the differences between chart types by aggregating all repetitions and trends (see \Cref{fig:s3_res_charts}).
There is clear evidence that \arousalCol{$C_{bar}$} is smaller than \arousalCol{$C_{sml}$}, \arousalCol{$C_{lin}$} and \arousalCol{$C_{sca}$}. 
There is also weak evidence that \arousalCol{$C_{sca} < C_{lin}$} and \arousalCol{$C_{sca} < C_{sml}$}. 
For valence, all CIs overlap with 0 -- there is only inconclusive indication that perhaps \valenceCol{$C_{lin} < C_{sml}$}, as the CI slightly overlaps with 0.

\inlinesection{Effect of trend}
To answer \textbf{H1}, we compare the effect of trend for all chart types combined (see \Cref{fig:s3_res_trends}).
There is strong evidence that \arousalCol{$T_{pos} > T_{neu}$} with a relatively large effect, and that \arousalCol{$T_{neu} > T_{neg}$}, but with a smaller effect.
There is strong evidence that \valenceCol{$T_{pos} > T_{neu}$} and that \valenceCol{$T_{neu} > T_{neg}$}, both with relatively large effects.
In sum, there is clear evidence that \arousalCol{$T_{pos} > T_{neu} > T_{neg}$} and \valenceCol{$T_{pos} > T_{neu} > T_{neg}$} by transitivity. 

\inlinesection{Difference between $T_{pos}$ and $T_{neu}$ for different chart types}
\label{sec:s3_res_2}
To answer \textbf{H2}, we compute the difference $T_{pos} - T_{neu}$ for each chart type and for each measurement after aggregating the repetitions. 
Then, we calculate and bootstrap the pairwise differences between chart types (see \Cref{fig:s3_res_posneu_neuneg}).
For arousal, the notable differences are that \arousalCol{$C_{lin} > C_{sca}$} and \arousalCol{$C_{lin} > C_{sml}$}; the evidence is weak and the effects are small.
For valence, the notable differences are that \valenceCol{$C_{bar} > C_{lin}$}, \valenceCol{$C_{bar} > C_{sca}$} and \valenceCol{$C_{bar} > C_{sml}$}; the evidence is relatively strong and the effects are small.

\inlinesection{Difference between $T_{neu}$ and $T_{neg}$ for different chart types}
To answer \textbf{H3}, we analyze the differences \arousalCol{$T_{neu} - T_{neg}$} and \valenceCol{$T_{neu} - T_{neg}$} for each chart type using the same method as for \textbf{H2} (see \Cref{fig:s3_res_posneu_neuneg}).
For arousal, the notable differences are that \arousalCol{$C_{lin} > C_{sca}$} and \arousalCol{$C_{lin} > C_{sml}$}; the evidence is weak and the effects are small.
For valence, no difference was found, as all CIs of differences largely overlap with 0.

\begin{figure}[!t]
\begin{minipage}[!t]{\columnwidth}
 \centering

   \includegraphics[alt={This figure shows the differences in arousal and valence ratings between line charts (C_lin) and smoothed line charts (C_sml) across different trends in Study 3. For arousal (pink bars), the overall difference is -0.02 [-0.29, 0.35], T_pos is -0.10 [-0.43, 0.29], T_neu is 0.09 [-0.21, 0.46], and T_neg is -0.05 [-0.35, 0.34]. For valence (blue bars), the overall difference is -0.32 [-0.61, 0.06], T_pos is -0.21 [-0.53, 0.19], T_neu is -0.31 [-0.63, 0.07], and T_neg is -0.45 [-0.78, -0.06]. Error bars represent the 95 percent confidence intervals for each condition.},width=.86\textwidth,keepaspectratio]{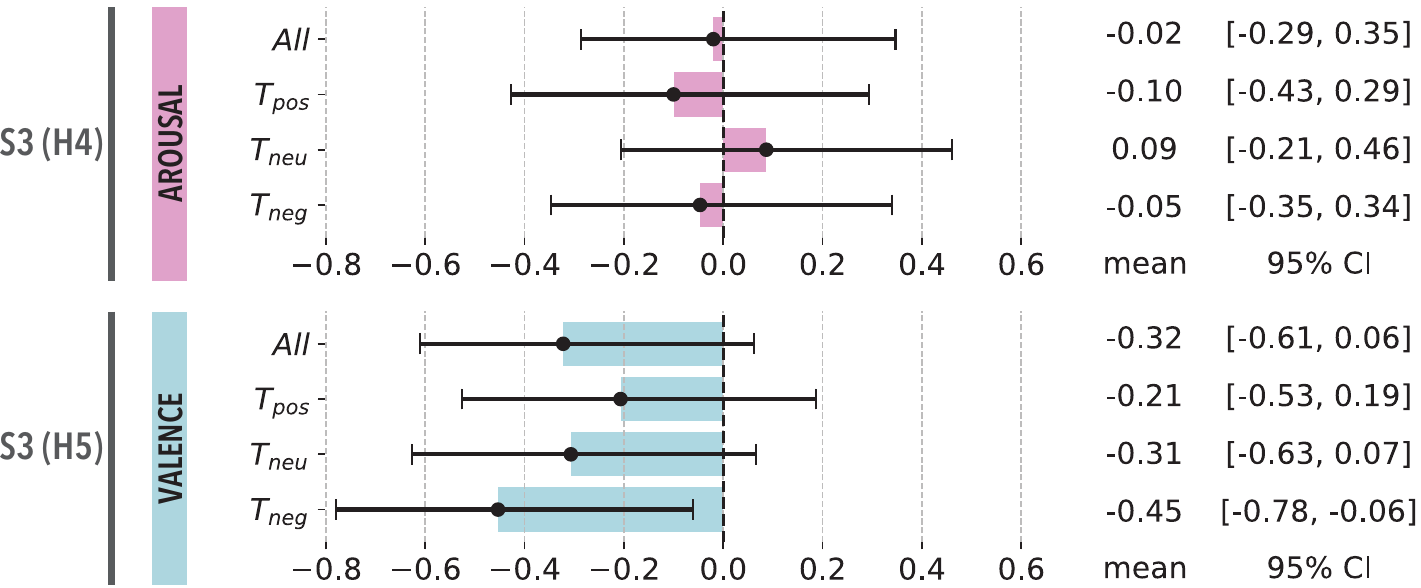}
  \captionof{figure}{S3: Differences between chart types of $C_{lin} - C_{sml}$ for each trend. \errorbardescrip}
  \label{fig:s3_res_jagsml}

  \vspace{1.5em}

   \includegraphics[alt={This figure shows the pairwise differences in standard deviations (σ) for trend and chart type in arousal and valence ratings in Study 3. The arousal difference (pink bar) has a mean of 0.54 with a 95 percent confidence interval of [0.20, 0.87]. The valence difference (blue bar) has a mean of 0.73 with a 95 percent confidence interval of [0.39, 1.06]. Error bars represent the 95 percent confidence intervals for each measurement.},width=.86\textwidth,keepaspectratio]{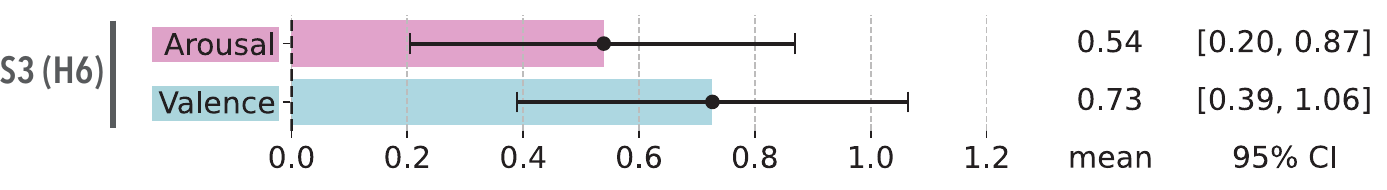}
  \captionof{figure}{S3: Pairwise difference between $\sigma_{trend}$ and $\sigma_{type}$. \errorbardescrip}
  \label{fig:s3_res_diff_trend_type}

  \vspace{1.5em}

     \includegraphics[alt={This figure displays participants' self-reported ratings on whether chart type or trend influences their choices more in Study 3. For arousal, 16 percent rated "mainly type," 8 percent rated 2, 28 percent rated "both equally," 10 percent rated 4, and 38 percent rated "mainly trend." For valence, 10 percent rated "mainly type," 12 percent rated 2, 32 percent rated "both equally," 4 percent rated 4, and 42 percent rated "mainly trend." Percentages reflect the distribution of responses.},width=1.0\textwidth,keepaspectratio]{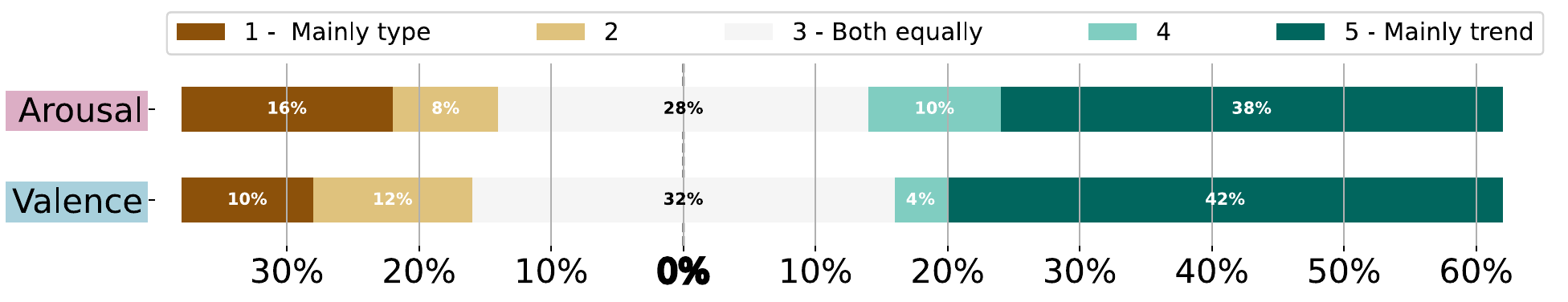}
  \captionof{figure}{S3: Participants' self-reported ratings on whether chart type or trend affect their choices more.}
  \label{fig:s3_res_likert}

  \vspace{1.5em}

    \includegraphics[alt={This figure presents nine conditions combining three variance levels (low, medium, high) and three chart types (scatterplot, jagged line chart, smoothed line chart) from Study 4. Each row represents a different chart type, and each column represents a different variance level. The x-axis for all charts is labeled "time." This set of charts visually demonstrates how different variance levels are represented across various chart types.},width=.62\columnwidth]{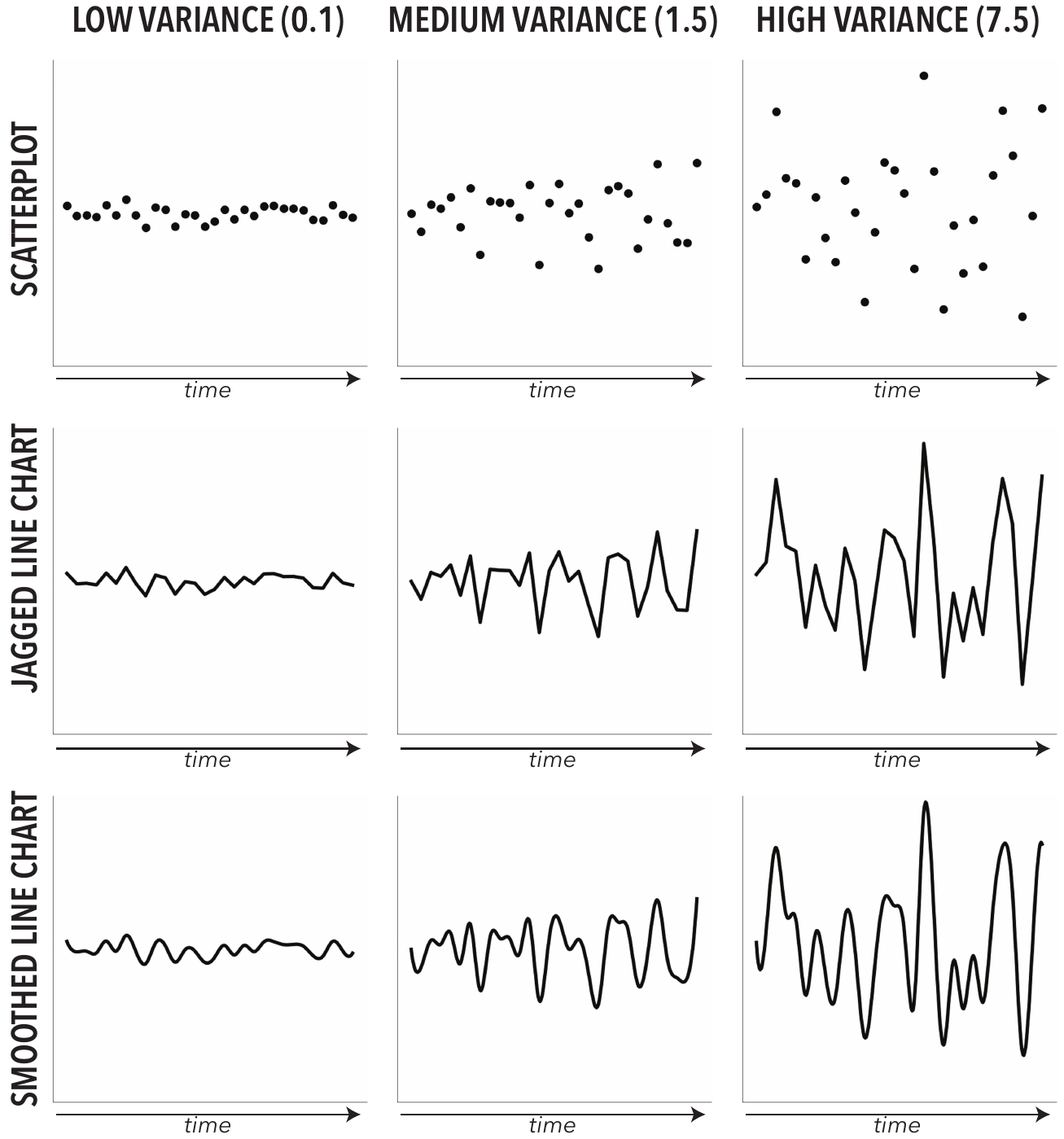}
  \caption{One set of charts for all 9 conditions used in S4 (three variance levels and three chart types), with a ``Time'' label on the $y$ axis.}
  \label{fig:s4_graphs}

  \vspace{-1.0em}
 \end{minipage}
\end{figure}

\inlinesection{Difference between jagged and smooth line charts}
To answer \textbf{H4} and \textbf{H5}, we analyze the differences \arousalCol{$C_{lin} - C_{sml}$} and \valenceCol{$C_{lin} - C_{sml}$} for each trend and for all trends combined (see \Cref{fig:s3_res_jagsml}).
For arousal (\textbf{H4}), no difference was found.
For valence (\textbf{H5}), there is evidence of a difference for $T_{neg}$, and some possible but inconclusive differences for $T_{neu}$ and all trends combined because the CIs slightly overlap with 0.

\inlinesection{Effect of trend and chart type}
To answer \textbf{H6}, we use the standard deviation ($\sigma$) like in S2 (see \Cref{fig:s3_res_diff_trend_type}).
There is strong evidence that \arousalCol{$\sigma_{trend} > \sigma_{type}$} and \valenceCol{$\sigma_{trend} > \sigma_{type}$}, and the differences are quite large.

\inlinesection{Self-reported perceptions}
Participants found that trend had a bigger influence on both measurements than chart type (see \Cref{fig:s3_res_likert}).

\subsection{S3 -- Discussion}

Results from S3 that \arousalCol{$T_{pos} > T_{neu} > T_{neg}$} and \valenceCol{$T_{pos} > T_{neu} > T_{neg}$} (\textbf{H1} confirmed) allow us to generalize results from S2 to other visualizations than scatterplots. 
Results also confirmed \textbf{H6}, with trend having a stronger effect than chart type on both measures.
Together, these results indicate that trend plays an important role in determining the emotion a visualization induces, even with no semantic meaning.

We formulated \textbf{H2} and \textbf{H3} because trend is more easily observed in line graphs and scatterplots than in bar charts~\cite{10.1145/3025453.3025922}, therefore we expected that the effect of trend on emotion would be more pronounced in these charts. 
However, \textbf{H2} is only partially supported, as \valenceCol{$T_{pos} - T_{neu}$} is smaller for bar charts and \valenceCol{$T_{neu} - T_{neg}$} is not significantly different for any of the chart types. 
Moreover, \textbf{H3} is rejected, as for \arousalCol{$T_{pos} - T_{neu}$} and \arousalCol{$T_{neu} - T_{neg}$} is not significantly different for any of the chart types.
Together, these findings indicate that positive trends have a stronger impact on valence in charts where the trend is easier to observe. 

Surprisingly, we found no evidence of \arousalCol{$C_{lin} > C_{sml}$} and only weak evidence of \valenceCol{$C_{lin} < C_{sml}$}, which means we do not replicate existing findings that smooth shapes are preferred to sharp shapes~\cite{aryani2020affective, bar2006humans, leder2011emotional}. 
We speculate this might be due to the data trend having a dominating effect on emotion compared to chart type, as shows our confirmation of \textbf{H6}.

\section{S4 -- Chart Types and Data Variance}
\label{sec:S4}

Given that we found trend has a strong effect on emotion in S2 and S3, there is reason to believe that other data characteristics could also affect emotion. 
In this fourth experiment, we study the effects of data variance (another data characteristic) and chart type on emotion.

\subsection{S4 -- Method}

\inlinesection{Design}
We determined three variance levels -- low ($V_{low}$), medium ($V_{med}$) and high ($V_{high}$) -- through visual inspection. 
We started by determining $V_{low}$ and $V_{high}$ so that they would produce graphs that are realistic while exhibiting clearly low variance for $V_{low}$, and while having a visibly erratic nature for $V_{high}$ (with minimum and maximum values nearing the $y$-scale's lower and upper bound). 
We then determined $V_{med}$ so that it would produce graphs with peaks and trough magnitudes between those of the graphs generated with $V_{low}$ and $V_{high}$. 
The resulting variance levels are $V_{low}: \sigma^2 = 0.1$ $(\sigma = 0.32)$, $V_{med}: \sigma^2 = 1.5$ $(\sigma = 1.22)$ and $V_{high}: \sigma^2 = 7.5$ $(\sigma = 2.74)$. 
To exhibit the variance in each graph, they all have the same $y$-scale with a range of $(-2,12)$.

We used a neutral trend ($\rho = 0$) for each graph to isolate the effects of variance and chart type. 
We studied three chart types: $C_{sca}$, $C_{sml}$ and $C_{lin}$ (we discarded $C_{bar}$ that is less relevant to studying variance).
With three levels of variance and three chart types, there were 9 conditions. 
With three repeated measures for each condition, there were three arousal blocks and three valence blocks. Each block had nine graphs -- one for each of the nine conditions. 
We used a different dataset for each repetition, and each condition showed the same datasets.

\inlinesection{Datasets}
We generated the datasets with different variance levels by first defining an array of 30 linearly spaced values ranging from 0 to 10 as the $x$-values.
Then, we defined an array of $y-values$, initially populated with the value 5 30 times. 
We added variance to each value in the $y-values$ array by drawing from a random normal distribution with a mean of 0 and a standard deviation equivalent to the square root of the target variance. 
This procedure was reiterated multiple times, and through rejection sampling, datasets of the target variance ($\pm 0.03$) and with a correlation coefficient of 0 ($\pm 0.03$) were selected. 
For each variance level, three datasets were generated, corresponding
to the three repetitions. 
\Cref{fig:s4_graphs} shows one set of 9 charts.

\inlinesection{Participants}
50 participants participated. 
They took on average 11 minutes 26 seconds to complete the study and were compensated \pounds 1.6.

\subsection{S4 -- Hypotheses}
We formulated five hypotheses for S4:
\begin{description}[noitemsep]
    \item[H1:] when aggregating over all chart types, \arousalCol{$V_{high} > V_{med} > V_{low}$} because uncertainty is known to induce anxiety~\cite{grupe2013uncertainty}, which is associated with high arousal~\cite{barrett1998discrete}.
    
    \item[H2:] when aggregating over all chart types, \valenceCol{$V_{high} < V_{med} < V_{low}$} because uncertainty is known to induce anxiety~\cite{grupe2013uncertainty}, which is associated with low valence~\cite{barrett1998discrete}.

    \item[H3:] \arousalCol{$C_{lin} > C_{sml}$}, as jagged shapes are associated with higher arousal than smooth shapes~\cite{bar2006humans, aryani2020affective}.
    
    \item[H4:] \valenceCol{$C_{lin} < C_{sml}$}, as jagged shapes are associated with lower valence than smooth shapes~\cite{bar2006humans} and S3 provided some evidence of this.

    \item[H5:] arousal will be higher for line charts than for scatterplot based on S3 results, with
    \arousalCol{$C_{sml} > C_{sca}$} (\textbf{H5a}) and \arousalCol{$C_{lin} > C_{sca}$} (\textbf{H5b}).

\end{description}

\subsection{S4 -- Results}
No responses were excluded. We gathered 2700 trial answers (50 participants $\times$ 3 chart types $\times$ 3 levels of variance $\times$ 2 measurements $\times$ 3 repetitions).

\inlinesection{Effect of variance}
To answer \textbf{H1} and \textbf{H2}, we compare the effect of variance for all chart types combined (see \Cref{fig:s4_res_variance_charts}).
There is strong evidence that \arousalCol{$V_{high} > V_{med}$} and that \arousalCol{$V_{med} > V_{low}$}, the latter with a larger effect (\textbf{H1} confirmed). 
There is also evidence that \valenceCol{$V_{high} < V_{med}$} but no difference was found between \valenceCol{$V_{med}$} and \valenceCol{$V_{low}$} (\textbf{H2} partially confirmed).

\inlinesection{Effect of smoothing}
To answer \textbf{H3} and \textbf{H4}, we analyze \arousalCol{$C_{sml} - C_{lin}$} and \valenceCol{$C_{sml} - C_{lin}$} (see \Cref{fig:s4_chart_differences}).
We found no evidence that \arousalCol{$C_{lin} > C_{sml}$} (\textbf{H3} rejected), but evidence that \valenceCol{$C_{lin} < C_{sml}$} (\textbf{H4} confirmed).

\inlinesection{Effect of chart type}
To answer \textbf{H5}, we analyze \arousalCol{$C_{sml} - C_{sca}$} and \arousalCol{$C_{lin} - C_{sca}$} (see \Cref{fig:s4_chart_differences}).
There is strong evidence that \arousalCol{$C_{sml} > C_{sca}$} (\textbf{H5a} confirmed) and \arousalCol{$C_{lin} > C_{sca}$} (\textbf{H5b} confirmed).

\subsection{S4 -- Discussion}

Results from S4 support \textbf{H1} and replicate previous findings from psychology that unpredictability induces anxiety~\cite{grupe2013uncertainty}, which is in turn associated with high arousal~\cite{barrett1998discrete}. 
Interestingly, the difference is larger for \arousalCol{$V_{med} - V_{low}$} than for \arousalCol{$V_{high} - V_{med}$}.
This suggests an effect with diminishing returns wherein arousal increases more quickly as variance increases from low to medium levels than from medium to high levels.

Our results only partially confirm \textbf{H2}.
Given the relationship between unpredictability and anxiety (which is associated with low valence~\cite{grupe2013uncertainty,barrett1998discrete}), our result that \valenceCol{$V_{high} < V_{med}$} was expected. 
However, we also expected to find \valenceCol{$V_{med} < V_{low}$}. 
This result, contrasted with the large difference for \arousalCol{$V_{med} - V_{low}$}, provides further evidence to support the independence of the arousal and valence dimensions of emotion.

Like in S3, we found no evidence of \arousalCol{$C_{lin} > C_{sml}$} (\textbf{H3} rejected), which means we do not replicate findings that smooth shapes are preferred to sharp shapes~\cite{aryani2020affective, bar2006humans, leder2011emotional}. 
We speculate this might be because the strong effect of variance hides the effect of smoothing on arousal.
On the other hand, we did find that \valenceCol{$C_{sml} > C_{lin}$} (\textbf{H4} confirmed), and this result, combined with S3's result that \valenceCol{$C_{sml} > C_{lin}$} for negative trends, shows that smoothing can be used to increase the valence of line charts.

Like in S3, we found strong evidence that both types of line charts we tested have higher arousal than scatterplots, confirming \textbf{H5}.


\section{S5 -- Chart Types and Data Density}
\label{sec:S5}
Given that both data trend and variance strongly impact arousal and valence, we extend our investigation to study the effects of data density (another data characteristic) and chart type on emotion.

\subsection{S5 -- Method}

\inlinesection{Design}
Similar to S4, we determined three density levels -- low ($D_{low}$), medium ($D_{med}$) and high ($D_{high}$) -- through visual inspection. 
We started by determining $D_{low}$ so that it would produce graphs that are realistic but not too simple. 
Then, we selected $D_{high}$ so that it would produce graphs that display as much data as possible while their peaks and troughs are still visually separable. 
We then determined $D_{med}$ so that it would produce graphs with a number of peaks and troughs between that of $D_{low}$ and $D_{high}$. 
The resulting density levels are $D_{low}$: 10 points, $D_{med}$: 20 points, and $D_{high}$: 100 points.
All datasets have a mean of 5 and all graphs have the same $y$-scale with a range of $(-2,12)$.

To isolate the effects of density and chart type, we used a neutral trend ($\rho = 0$) for each graph. 
We again studied $C_{sca}$, $C_{sml}$ and $C_{lin}$ because these chart types are commonly used in scenarios with varying degrees of data density.
With 3 levels of density and 3 chart types, there were 9 conditions. 
With 3 repeated measures for each condition, there were three arousal blocks and three valence blocks. 
Each block had 9 graphs -- one for each of the nine conditions. 
We used a different dataset for each repetition, and each condition showed the same datasets.

\begin{figure}[!t]
\begin{minipage}[!t]{\columnwidth}
 \centering
    
  \includegraphics[alt={This figure displays the differences in arousal and valence ratings between different variance levels across all chart types in Study 4. For arousal (pink bars), the comparisons show the mean differences and 95 percent confidence intervals: V_high - V_med (0.86 [0.57, 1.17]) and V_med - V_low (1.94 [1.45, 2.45]). For valence (blue bars), the differences are: V_high - V_med (-0.42 [-0.74, -0.10]) and V_med - V_low (-0.22 [-0.67, 0.20]). Error bars represent the 95 percent confidence intervals for each condition.},width=.88\textwidth,keepaspectratio]{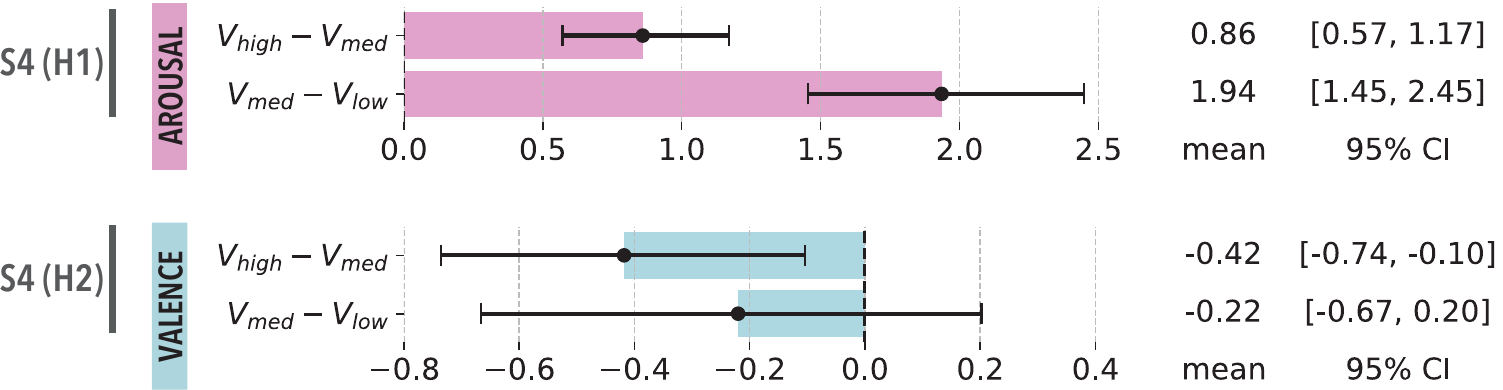}
  \captionof{figure}{S4: Differences between variance levels (all chart types). \errorbardescrip}
  \label{fig:s4_res_variance_charts}

  \vspace{1.25em}

  \includegraphics[alt={This figure shows the differences in arousal and valence ratings between different chart types across all variance levels in Study 4. For arousal (pink bars), the comparisons show the mean differences and 95 percent confidence intervals: smoothed line versus scatterplot (0.79 [0.44, 1.14]), line versus scatterplot (0.73 [0.40, 1.04]), and smoothed line versus line (0.06 [-0.08, 0.19]). For valence (blue bar), the difference between the smoothed line and the line is 0.30 [0.10, 0.48]. Error bars represent the 95 percent confidence intervals for each condition.},width=.88\textwidth,keepaspectratio]{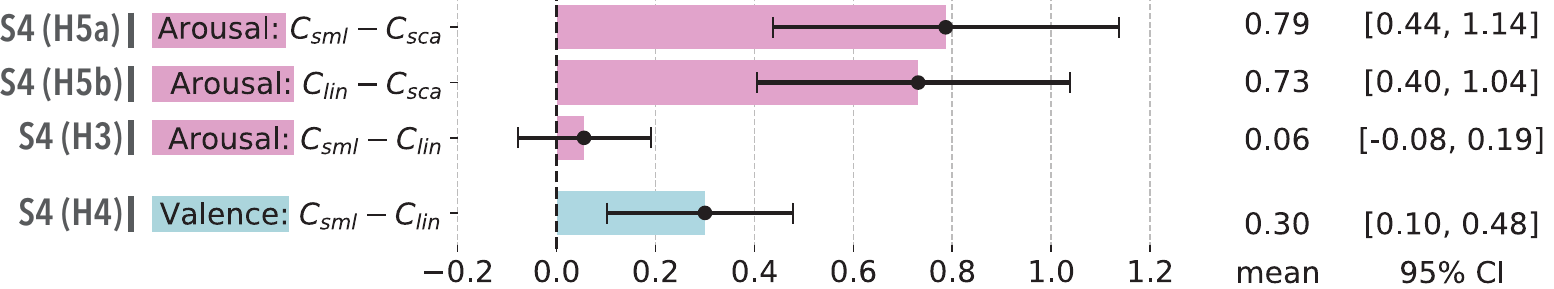}
  \captionof{figure}{S4: Differences between chart types (all variance levels). \errorbardescrip}
  \label{fig:s4_chart_differences}

  \vspace{1.25em}

    \includegraphics[alt={This figure presents nine conditions combining three density levels (low, medium, high) and three chart types (scatterplot, jagged line chart, smoothed line chart) from Study 5. Each row represents a different chart type, and each column represents a different density level. The x-axis for all charts is labeled "time." This set of charts visually demonstrates how different density levels are represented across various chart types.},width=.65\columnwidth]{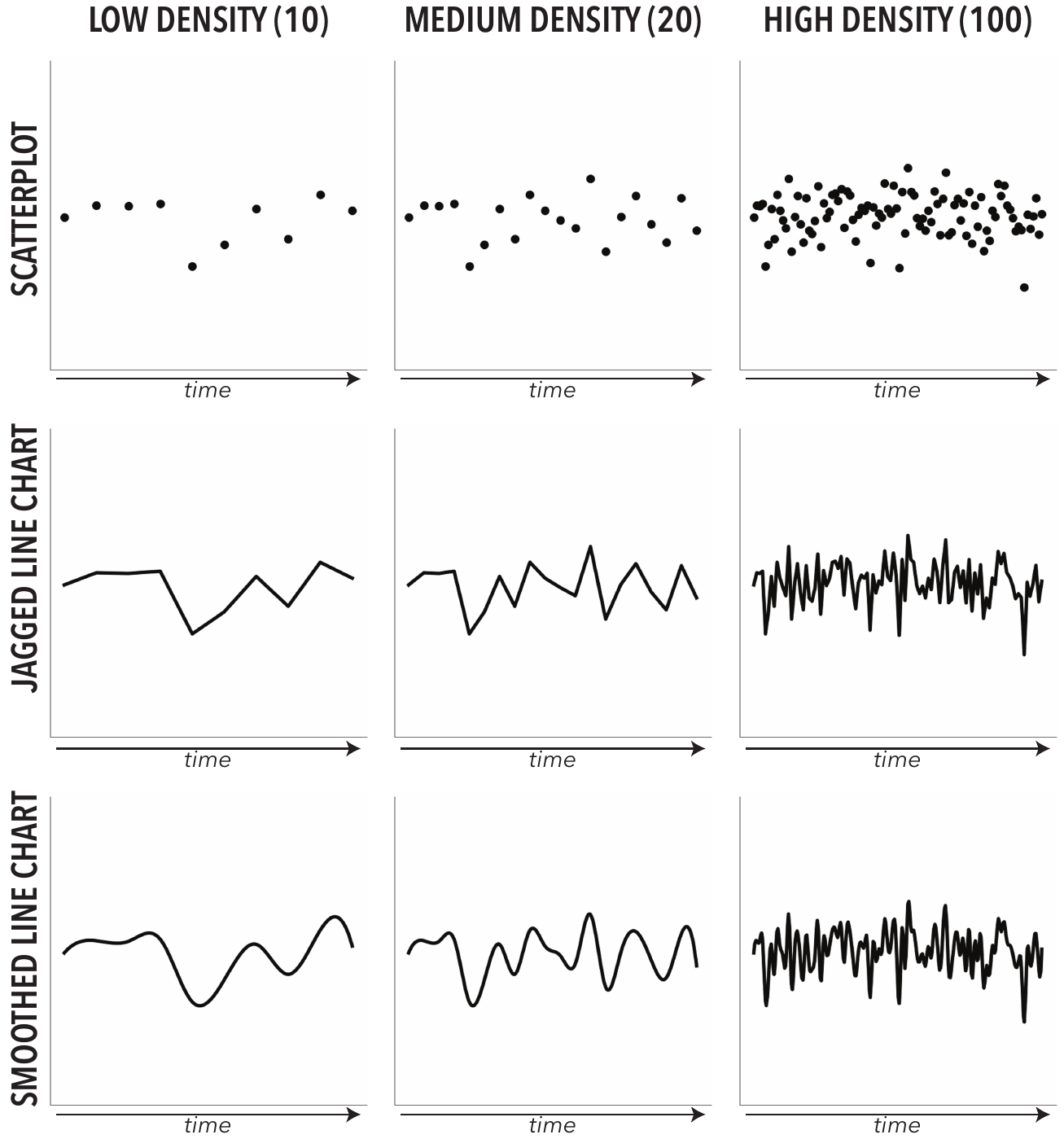}
  \caption{One set of charts for all 9 conditions used in S5 (three density levels and three chart types), with a ``Time'' label on the $y$ axis.}
  \label{fig:s5_graphs}

 \end{minipage}
 \vspace{-1.5em} 
\end{figure}

\inlinesection{Datasets}
We generated datasets with different density levels by first defining arrays of 10, 20 and 100 linearly spaced values on the $x-$axis. 
Then, we drew a $y$ value from a random normal distribution ($\mu = 5$, $\sigma = 1$) for each $x$ value. 
We repeated this process to select datasets that had similar variance ($\pm 0.03$) and a correlation coefficient near 0 ($\pm 0.03$). 
For each density level, three datasets were generated, corresponding to the three repetitions. 
\Cref{fig:s5_graphs} shows one set of 9 charts.

\inlinesection{Participants}
50 participants participated. 
They took on average 10 minutes 19 seconds to complete the study and were compensated \pounds 1.6. 

\subsection{S5 -- Hypotheses}
We formulated six hypotheses for S5:
\begin{description}[noitemsep]
    \item[H1:] when aggregating over all chart types, \arousalCol{$D_{high} > D_{med} > D_{low}$} 
    because both increased cognitive load~\cite{macpherson2017acoustic} and increased task complexity~\cite{malmberg2022exploring}, that density should affect, result in increased arousal.
    
    \item[H2:] for $C_{sca}$, \valenceCol{$D_{low} < D_{med} < D_{high}$} because we know that uncertainty is associated with low valence~\cite{grupe2013uncertainty} and higher density should make the flat trend more discernible and less uncertain~\cite{lauer1989density}.
    
    \item[H3:] for $C_{sml}$ and $C_{lin}$, \valenceCol{$D_{low} > D_{med} > D_{high}$} because uncertainty is associated with low valence~\cite{grupe2013uncertainty} and higher density 
    should make the charts appear more chaotic and more uncertain.

    \item[H4:] \arousalCol{$C_{sml}$ < $C_{lin}$}, like in S3 and S4.

    \item[H5:] \valenceCol{$C_{sml} > C_{lin}$}, like in S3 and S4.

    \item[H6:] arousal will be higher for line charts than for scatterplot like in S4, with
    \arousalCol{$C_{sml} > C_{sca}$} (\textbf{H6a}) and \arousalCol{$C_{lin} > C_{sca}$} (\textbf{H6b}).
    
\end{description}

\subsection{S5 -- Results}
No responses were excluded. In total, we gathered 2700 trial answers (50 participants $\times$ 3 chart types $\times$ 3 levels of density $\times$ 2 measurements $\times$ 3 repetitions).

\begin{figure}[!t]
\begin{minipage}[!t]{1.0\columnwidth}
 \centering
 
  \includegraphics[alt={This figure shows the differences in arousal and valence ratings between different density levels across all chart types in Study 5. For arousal (pink bars), the comparisons show the mean differences and 95 percent confidence intervals: D_high - D_med (1.14 [0.72, 1.57]) and D_med - D_low (0.55 [0.32, 0.80]). For valence (blue bars), the differences are: D_high - D_med (-1.08 [-1.43, -0.72]) and D_med - D_low (-0.45 [-0.67, -0.23]). Error bars represent the 95 percent confidence intervals for each condition.}, width=.88\textwidth]{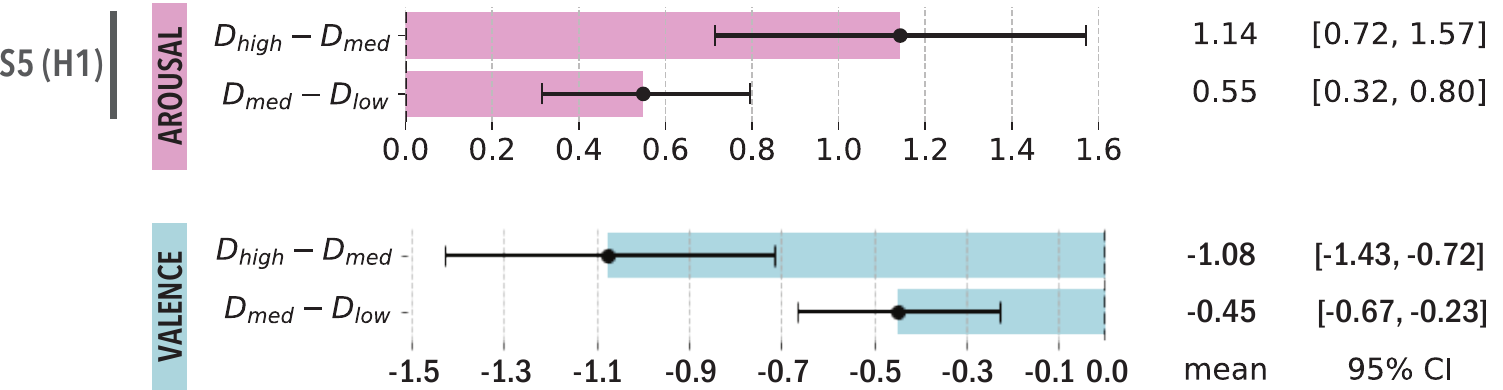}
  \captionof{figure}{S5: Differences between density levels (all chart types). \errorbardescrip}
  \label{fig:s5_res_density_charts}

  \vspace{1.5em}

  \includegraphics[alt={This figure shows the differences in valence ratings between different density levels for each chart type in Study 5. For scatterplots, the valence difference between high and medium density is -1.07 with a 95 percent confidence interval of [-1.46, -0.69], and between medium and low density is -0.45 [-0.73, -0.16]. For jagged line charts, the difference between high and medium density is -0.87 [-1.25, -0.47], and between medium and low density is -0.26 [-0.53, 0.00]. For smoothed line charts, the difference between high and medium density is -1.29 [-1.76, -0.81], and between medium and low density is -0.63 [-0.93, -0.32]. Error bars represent the 95 percent confidence intervals for each condition.}, width=.88\textwidth]{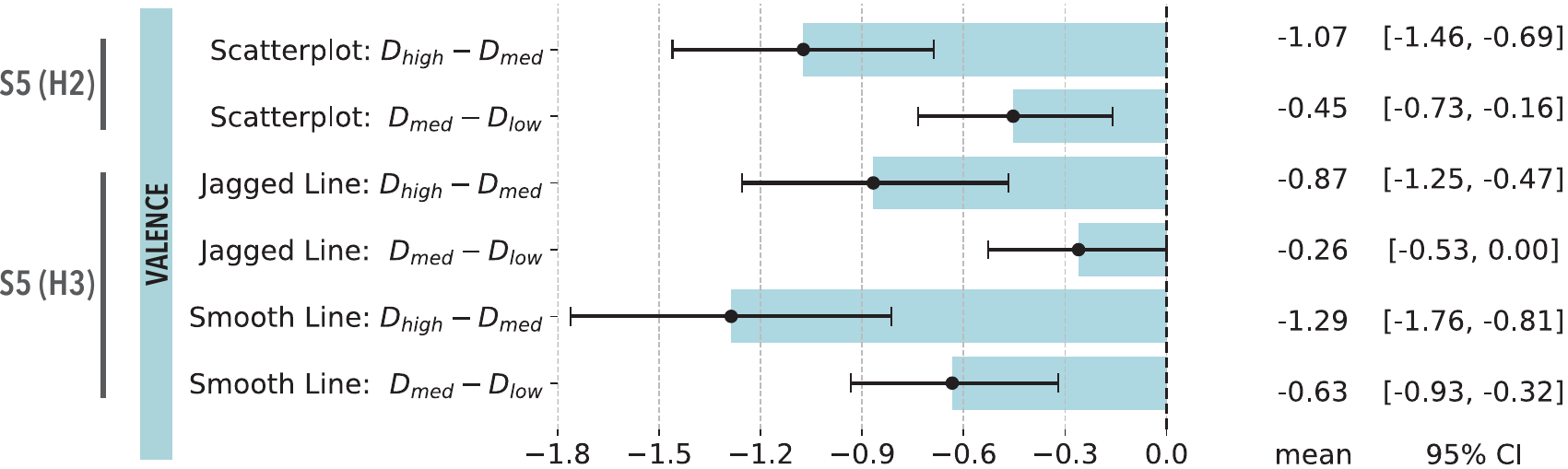}
  \captionof{figure}{S5: Differences in valence between density levels for each chart type.}
  \label{fig:s5_res_density_valence}

  \vspace{0.5em}

    \includegraphics[alt={This figure shows the differences in arousal and valence ratings between different chart types across all density levels in Study 5. For valence (blue bar), the difference between smoothed line and line charts is 0.47 with a 95 percent confidence interval of [0.27, 0.67]. For arousal (pink bars), the comparisons show the mean differences and 95 percent confidence intervals: smoothed line versus line charts (-0.21 [-0.40, -0.00]), line versus scatterplot (0.53 [0.25, 0.82]), and smoothed line versus scatterplot (0.32 [0.04, 0.59]). Error bars represent the 95 percent confidence intervals for each condition.},width=.88\textwidth]{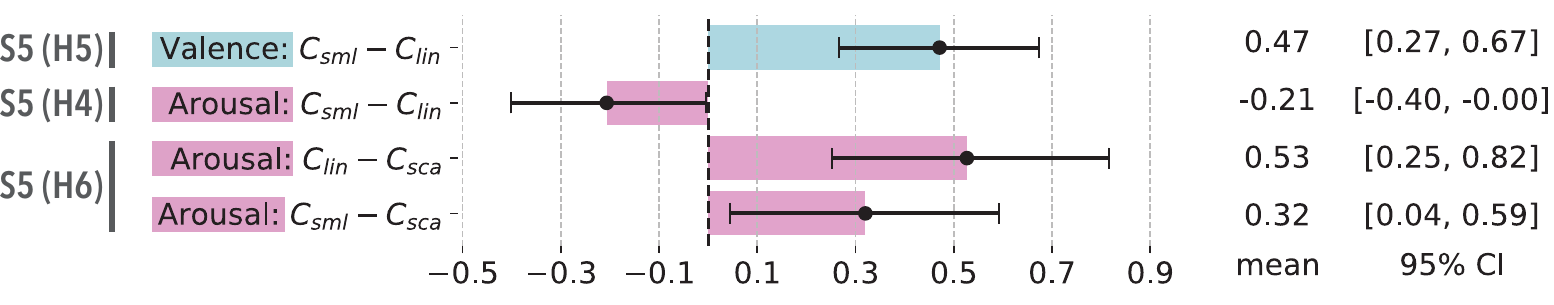}
  \captionof{figure}{S5: Differences between chart types (all density levels). \errorbardescrip}
  \label{fig:s5_chart_differences}
  
 \end{minipage}
 \vspace{-1em} 
\end{figure}

\inlinesection{Effect of Density}
To answer \textbf{H1}, we compare the effect of density on arousal for all chart types combined (see \Cref{fig:s5_res_density_charts}). 
There is evidence that \arousalCol{$D_{high} > D_{med}$} and \arousalCol{$D_{med} > D_{low}$}, the latter with a smaller effect (\textbf{H1} confirmed).
To answer \textbf{H2} and \textbf{H3}, we look at each chart type (see \Cref{fig:s5_res_density_valence}), and for all find evidence that \valenceCol{$D_{high} > D_{med}$} and \valenceCol{$D_{med} > D_{low}$}, the latter with smaller effects (\textbf{H2} rejected, \textbf{H3} confirmed).

\inlinesection{Effect of Smoothing} 
To answer \textbf{H4} and \textbf{H5}, we analyze \arousalCol{$C_{sml} - C_{lin}$} and \valenceCol{$C_{sml} - C_{lin}$} (see \Cref{fig:s5_chart_differences}).
There is some evidence that \arousalCol{$C_{sml} < C_{lin}$} but with a small effect and a confidence bound close to 0 (\textbf{H4} weakly confirmed).
There is strong evidence that \valenceCol{$C_{sml} > C_{lin}$} (\textbf{H5} confirmed).

\inlinesection{Effect of Chart Type} 
To answer \textbf{H6}, we analyze \arousalCol{$C_{sml} - C_{sca}$} and \arousalCol{$C_{lin} - C_{sca}$} (see \Cref{fig:s5_chart_differences}).
There is evidence that \arousalCol{$C_{sml} > C_{sca}$} (\textbf{H6a} weakly confirmed) and \arousalCol{$C_{lin} > C_{sca}$} (\textbf{H6b} confirmed).

\subsection{S5 -- Discussion}
Results from S5 support \textbf{H1} (higher density leads to higher arousal) and replicate previous findings that high cognitive load, task complexity, and uncertainty are associated with high arousal~\cite{macpherson2017acoustic, grupe2013uncertainty, malmberg2022exploring} as increasing data density increases the amount of information to process. 

We hypothesized that increasing density in scatterplots would increase valence because humans are averse to uncertainty~\cite{grupe2013uncertainty}, and a more obvious flat trend~\cite{lauer1989density} would induce less uncertainty. However, our results rejected \textbf{H2}. 
Perhaps the absence of meaning in the data or of visualization task to perform led to low participants' interest in understanding the data or in estimating data characteristics, making uncertainty and its emotional associations less relevant. 
It might also be that the relationship between varying levels of density and perceived uncertainty in scatterplots is not that clear -- this calls for further research. 
On the other hand, our hypothesis that line charts with high density induce low valence due to their chaotic nature (\textbf{H3}) was confirmed. 

Results from S5 complement those from S3 and S4 regarding the Kiki-Bouba effect (that rounded stimuli have lower arousal and higher valence than sharp stimuli~\cite{bar2006humans, aryani2020affective}).
While in S3 and S4 we did not find evidence that \arousalCol{$C_{sml} < C_{lin}$}, we did find such evidence in S5 and confirmed \textbf{H4}.
Similarly, we had found some evidence that \valenceCol{$C_{sml} > C_{lin}$} (only with a negative data trend in S3, and when aggregating all variance levels in S4), and we did find further evidence of this in S5 and confirmed \textbf{H5}.

Last, like in S3 and S4, we found that both types of line charts we tested have higher arousal than scatterplots, confirming \textbf{H6}.

\section{What this Means for Visualization Design}

\add{These} five studies provide the first systematic exploration of the effects of several immutable data characteristics and visual features on emotion. 
All the factors we tested do affect emotion to a certain degree, even though data had no semantic meaning. 
Using \Cref{fig:summary-results}, we summarize our findings and discuss implications for considering emotion when designing visualizations before discussing limitations and future work.

\subsection{Summary of findings}

\inlinesection{Reflection on Method}
Reflecting on the method of using SAM and crowdsourcing experiments, the results are consistent between different studies and with previous work (\eg~\cite{Bartram:2017:ACV}). This helps us confirm the validity of the study method and the meaningfulness of the results.

\inlinesection{Effect of color}
Our results replicate findings that the choice of color affects emotion~\cite{Bartram:2017:ACV}; they further show that the effect of color is the stronger of the ones we studied: color has a stronger effect than trend, which in turn has a stronger effect than chart type.
Although the grey palette is not emotionally neutral, this palette elicited the lowest arousal and valence among all conditions in S1 and S2 and exhibited the smallest variation between S1 and S2. 
Since the grey palette is both stable and on the low end of the arousal/valence scale, it can be used as a baseline to measure the effect of other color palettes on emotion.

\inlinesection{Effect of data characteristics}
\textit{Data trend} unsurprisingly has an effect on emotion: positive trends induce higher valence and arousal than neutral trends, that in turn induce higher valence and arousal than negative trends (from S2 and S3). 
\textit{Data variance} also has an effect: higher variance leads to higher arousal and lower valence; in particular, arousal ratings are more sensitive at the low end of the variance scale, with diminishing sensitivity as variance increases. 
\textit{Data density} also has an effect: higher density leads to higher arousal and lower valence.
This means that data trend, variance and density should be taken into account when designing visualizations aimed at conveying certain emotions. 

\inlinesection{Effect of chart type}
We did not formulate hypotheses on the ordering of chart types in terms of induced arousal and valence due to the lack of previous research on that topic. 
However, exploratory analysis revealed that line charts (both smooth and jagged) induce higher arousal than scatterplots, which in turn induce higher arousal than bar charts. 
We also found no evidence that jagged line charts, bar charts and scatterplots would induce different valence; and that the impact of a positive trend on valence (relative to a neutral trend) is higher in chart types which better \add{show} trends. 
This suggests that chart type can affect the perceived valence when the data has a positive trend. 
In addition, we found \add{some} 
evidence that smoothed line charts induce higher valence than jagged line charts, which corresponds to the known preference for curved objects over sharp ones~\cite{bar2006humans}.
Put together, these findings indicate that chart type has an impact on the emotional response of viewers, both independently and in conjunction with the data trend.

\subsection{Information Visualization: Emotion for Design}
We have found that data trend, data variance, data density, color palette and chart type all have a certain effect on emotion. Below we discuss what these effects mean for visualization design and provide \guidelineCol{practical suggestions for designing visualizations while considering emotion.}

\inlinesection{Considering Immutable Data Characteristics}
Designers have little control over data characteristics such as trend, variance, and density, that do impact emotions. 
However, designers can change the emotional impact of these characteristics through visualization design choices. 

For example, altering the way a trend is perceived will change its impact on arousal and valence. 
Designers can \guidelineCol{employ logarithmic scales or adjust axis start and end values to amplify or diminish trend perception.} For instance, shortening the range of the vertical axis in a visualization with a positive data trend would diminish the perception of that trend and counterbalance the heightened valence and arousal induced by a positive trend.
Similar techniques can be applied to data variance and data density. 
For instance, designers can \guidelineCol{increase the range of values of the $y$ axis to make that data appear as if it has lower variance} -- leading to lower arousal and higher valence. 
Since high data density is associated with high arousal and low valence, designers could also \guidelineCol{make a visualization larger to make the data appear as if it is less dense} -- leading to lower arousal and higher valence.

\inlinesection{Considering Chart Type}
Many guidelines and recommendations exist to choose a chart type, informed by various factors other than emotion (\eg data types, audience, tasks to support, or implementation constraints). 
However, our results provide evidence that emotion is yet another aspect that designers should consider when selecting a type of chart, as it can influence the induced emotion, and it can be used to alter the emotional impact of the data trend. 
For instance, a designer may want to \guidelineCol{avoid bar charts if heightening induced valence is a goal.}
That is because we found a less pronounced difference in valence ratings for positive and neutral trends in bar charts compared to other chart types. 
This observation echoes findings that trend is less discernible in bar charts~\cite{10.1145/3025453.3025922}, as a greater emotional impact from data trend would be expected when trends are more easily perceived. 
We also observed that bar charts evoke lower arousal than all other tested chart types, while line charts produce higher arousal than all other tested chart types.
As a result, a designer should \guidelineCol{prefer line graphs over scatterplots and scatterplots over bar charts if heightening induced arousal is a goal.}

S3, S4 and S5 compared 
smoothed and jagged line charts, thus providing data about the kiki-bouba effect that is characterized by a preference for rounded objects over sharp ones. 
We found that smoothed line charts (with round features) induce higher valence than jagged line charts (with sharp features), especially when there is a negative data trend.
This means that designers can \guidelineCol{smoothen a line chart to increase its induced valence without altering its induced arousal}. 

\inlinesection{Considering Color}
Choosing an appropriate color palette might be the best way for designers to induce 
emotions. 
Color is the most impactful factor on emotion among the ones we tested.
This finding from the perspective of emotion corroborates previous work that arrived at similar conclusions from the perspective of perception.
In Bertin's own (translated) words, \textit{``[color] will always remain the privileged variable of the graphic designer in problems involving differentiation [as color has] an undeniable psychological attraction''}~\cite[p.~91]{bertin2010semiology}.
This claim has since then been demonstrated in psychology, where it has been shown that color is more conducive to fast visual search than other encodings such as line orientation~\cite{nothdurft1993role}; and in visualization, where there is strong evidence of the power of color, in particular for detection tasks~\cite{healey-preattentive} and for representing nominal data dimensions~\cite{bertin2010semiology,maceachren2004maps,mackinlay-automating}.
Our recommendation is, therefore, to \guidelineCol{consider color at the end of a visualization design process and as a way to inhibit, balance, or strengthen an emotion based on the data characteristics and based on what other visualization features have been used} -- a guideline often suggested based on experience and established visualization guidelines that is given additional importance here, from the perspective of emotion.

Considering the color palette last in visualization design allows designers to consider the emotion induced by other factors (\eg trend, density, variance, chart type) and to counter or amplify the effect of these factors by using a color palette that induces opposite or similar emotion, respectively. 
For example, to represent a dataset about an economic crisis with a negative trend in an emotionally neutral way, one could opt for an exciting color palette that would counterbalance the negative emotional response induced by the negative trend.

\add{
\inlinesection{Possible Side Effects}
Making design decisions to impact viewers' emotions does have side effects.
For example, positive emotional priming can increase visual judgment performance~\cite{harrisonAffectivePriming}. 
Our results suggest that techniques like scaling an axis to exaggerate trends could increase the perceived valence of a visualization and act as a sort of positive visual priming which could lead to improved visual judgment.
However, we also know that the distortion induced by the alteration of the axis can significantly alter the perceived message of a visualization~\cite{pandeyDeceptive}, and could come at the cost of making comparisons of values more difficult.
Sometimes, making such adjustments is difficult. For example, a designer might want to make a visualization with high density larger to increase its induced valence, but they might not have control over the display size. 
Therefore, designers should consider emotion together with the many other dimensions of visualization design.
}

\subsection{Limitations and Future Work}
Like with any controlled experiment, we had to make study design choices. 
While these choices allow for control and precision, they also result in limitations in terms of generalizability~\cite{mcgrath1995methodology}.

One limitation of our studies is that the \textit{data trends} 
were all roughly linear with the same coefficients.
We wonder to what extent the form or the intensity of trends could influence the results \add{(e.g., an exponential trend might more strongly affect emotion than a linear trend)}. 
Studying trend variations is a promising avenue for future work.

Our results suggest that there might be an interesting non-linear relationship between \textit{data variance} and arousal. 
Since our experiment was limited to three levels of data variance, we do not have enough granularity to delve deeper into this relationship. 
Future work could model this relationship to get a clearer picture than we could offer.

\add{In our studies, we only varied the horizontal \textit{data density}.}
It is possible that vertical data density has a different relationship with emotion than horizontal density. 
Additionally, we were unable to find a satisfactory explanation for why the valence of scatterplots was inversely related to density. 
Future work should look into the relationship between data density in scatterplots, correlation perception, and emotion.

We did not consider the accuracy of \textit{information extraction}, which is an essential criterion for designing good visualizations. 
For example, although the smoothed line charts we used in S3, S4, and S5 display the same data as their non-smoothed version, smoothing creates visual artifacts that are not 
true to the data.
Such modifications (e.g., smoothing, scattering, blurring, aggregating) likely affect the readability of information. 
Similar concerns apply to the size of a stimulus: the accuracy and efficiency of information retrieval depend on the size of visual idioms; designs that are too small or too large can be hard to read or too cluttered. 
Future work could look into these aspects by considering both emotional response and task-based performance metrics. 

\add{In S3, we controlled for ink ratio, which made the bars in the bar chart thinner than the elements in other chart types. This reduced prominence of the bars may contribute to the lower perceived arousal ratings for bar charts. Future work should compare the effects of ink ratio and feature prominence on induced emotion.}

\add{We did not consider the potential impact of culture on emotions. Most study participants were English-speaking from Western countries (see supplemental material), and cultural differences in color meanings~\cite{coloursincultures} and reading direction~\cite{naturereadingdirection} likely alter visual interpretation. Future work should investigate the role culture plays in this context.}

Finally, we 
excluded \textit{semantics} 
\add{to} isolate the effects of experimental factors, like others have done before (e.g.~\cite{Bartram:2017:ACV}). 
However, semantics surely affect a viewer's emotion~\cite{Lan:2021:SSL, kennedy2018feeling}, and real-life charts do have semantics. 
Researchers have argued that data storytelling with visualization 
\add{could} make viewers emotionally connect with the story~\cite{Gershon2001}; and work related to emotion has studied, for example, how storytelling can persuade viewers~\cite{pandey-persuasion} or make them empathize with a topic~\cite{liem-empathy,boy-Anthropomorphism}. 
Investigating the relationship between semantics and 
\add{visualization features} is promising, especially 
\add{in a storytelling context.}


\section{Conclusions}
We conducted five studies investigating the relationship between emotion and color, chart type, data trend, data variance, and data density. 
We found that each has an impact on the emotion induced by a visualization, even if the data has no semantic meaning. 

In Study 1, we replicated results from previous work using the SAM rating scales.
Then, through Study 2, we quantified the effect of trend on emotion, which we confirmed in Study 3. 
In Study 3, we also found that chart type has a certain influence on emotion. 
In Study 4, we investigated the effects of chart type and data variance, confirming previous results about chart type and uncovering the emotional effect of data variance. 
In Study 5, we tested chart type and data density and quantified the emotional impact of data density.

Put together, these results provide an empirical foundation and practical considerations that designers can use to create charts that account for the emotional impact that visual features of visualization have on viewers. They also pave the way for future research in identifying and quantifying what aspects of a visualization affect emotion.







\acknowledgments{%
This work was supported by NSERC USRA (\#521850) and NSERC Discovery Grant (RGPIN-2019-05422).}

\add{
\section{Supplemental Materials}
The supplemental materials, available at \url{https://osf.io/ywjs4/}, include the following:
\begin{itemize}[noitemsep]
    \item \texttt{00-study\_material.pdf}: sample pre-block study instructions, and sample valence and arousal stimuli with rating interface.
    \item \texttt{01-demographic-data.pdf}: figure summarizing participant demographics for each study.
    \item \texttt{02-data\_column\_legend.txt}: description of the columns in the five raw data files.
    \item \texttt{S1-data.csv} to \texttt{S5-data.csv}: raw data from each study.
    \item \texttt{S1-analysis.ipynb} to \texttt{S5-analysis.ipynb}: analysis script for each study.    
\end{itemize}
}

\bibliographystyle{abbrv-doi-hyperref-narrow}

\bibliography{references}




\end{document}